\documentclass[twocolumn,superscriptaddress,aps,preprintnumbers,amsmath,amssymb,prl,nofootinbib]{revtex4}
\usepackage{epsf}
\usepackage{amsmath,amssymb}
\input{colordvi.tex}
\usepackage[usenames,dvipsnames]{color}
\usepackage{graphicx}
\usepackage{ascmac}
\usepackage{hyperref}
\usepackage{tikz}
\usepackage{tikz-feynhand}

\begin{document}

\preprint{KEK-QUP-2024-0006, TU-1226, KEK-TH-2607, KEK-Cosmo-0341, IPMU24-0007}

\title{Gravitational Wave Probe of Planck-scale Physics After Inflation}

\author{Weiyu Hu\footnote{weiyuhu@stu.pku.edu.cn}}
\affiliation{School of Physics, Peking University, Beijing 100871, China}

\author{Kazunori Nakayama\footnote{kazunori.nakayama.d3@tohoku.ac.jp}}
\affiliation{Department of Physics, Tohoku University, Sendai, Miyagi 980-8578, Japan}
\affiliation{International Center for Quantum-field Measurement Systems for Studies of the Universe and Particles (QUP), High Energy Accelerator Research Organization (KEK), 1-1 Oho, Tsukuba, Ibaraki 305-0801, Japan}

\author{Volodymyr Takhistov\footnote{vtakhist@post.kek.jp}} 
\affiliation{International Center for Quantum-field Measurement Systems for Studies of the Universe and Particles (QUP), High Energy Accelerator Research Organization (KEK), 1-1 Oho, Tsukuba, Ibaraki 305-0801, Japan}
\affiliation{Theory Center, Institute of Particle and Nuclear Studies (IPNS), High Energy Accelerator Research Organization (KEK), Tsukuba 305-0801, Japan
}
\affiliation{Graduate University for Advanced Studies (SOKENDAI), \\
1-1 Oho, Tsukuba, Ibaraki 305-0801, Japan}
\affiliation{Kavli Institute for the Physics and Mathematics of the Universe (WPI), UTIAS, \\The University of Tokyo, Kashiwa, Chiba 277-8583, Japan}

\author{Yong Tang\footnote{tangy@ucas.ac.cn}}
\affiliation{School of Astronomy and Space Science,
University of Chinese Academy of Sciences (UCAS), Beijing 100049, China}
\affiliation{School of Fundamental Physics and Mathematical Sciences,
Hangzhou Institute for Advanced Study, UCAS, Hangzhou 310024, China}
\affiliation{International Center for Theoretical Physics Asia-Pacific, Beijing 100049, China}

\begin{abstract}

Particle decays are always accompanied by the emission of graviton quanta of gravity through bremsstrahlung processes. However, the corresponding branching ratio is suppressed by the square of the ratio of particle's mass to the Planck scale. The resulting present abundance of gravitational waves (GWs), composed of gravitons, is analogously suppressed. We show that superheavy particles, as heavy as the Planck scale, can be naturally produced during the post-inflationary reheating stage in the early Universe and their decays yield dramatic amounts of GWs over broad frequency range. GW observations could hence directly probe Planck-scale physics, notoriously challenging to explore.

\end{abstract}

\maketitle

\paragraph{Introduction.}-- 
Discovery of GWs, composed of graviton quanta of gravity,  from binary mergers by LIGO and Virgo~\cite{LIGOScientific:2016aoc,LIGOScientific:2017vwq}
as well as recent evidence of the first detection of stochastic GW background reported by pulsar timing array collaborations~\cite{NANOGrav:2023gor,Xu:2023wog,Reardon:2023gzh,EPTA:2023fyk} consolidated GWs as prime instruments for probing fundamental physics. Cosmological GWs
can offer unique insights into high energy phenomena appearing in the extreme environments of the early Universe, but not accessible to conventional laboratories. Among notable cosmological GW sources are primordial (or inflationary) perturbations, formation of topological defects, first-order phase transitions and preheating processes (see e.g.~\cite{Maggiore:2018sht} for review). 

Recently, a broad novel class of cosmological GW sources associated with {\it quantum} gravitational processes have been put forth: 
scattering of particles in thermal plasma~\cite{Ghiglieri:2015nfa,Ghiglieri:2020mhm,Ringwald:2020ist}, gravitational bremsstrahlung from inflaton/particle decay~\cite{Nakayama:2018ptw,Huang:2019lgd,Barman:2023ymn,Bernal:2023wus}, inflaton/particle annihilation to gravitons~\cite{Ema:2015dka,Ema:2016hlw,Ema:2020ggo,Ghiglieri:2022rfp,Ghiglieri:2024ghm,Choi:2024ilx}, direct inflaton decay to gravitons~\cite{Ema:2021fdz,Mudrunka:2023wxy,Tokareva:2023mrt}.
Here, we make distinction with GWs stemming from classical as well non-perturbative processes such as inflaton fragmentation into solitonic oscillons~\cite{Zhou:2013tsa,Antusch:2016con,Liu:2017hua,Lozanov:2019ylm,Amin:2018xfe,Hiramatsu:2020obh,Lozanov:2023aez,Lozanov:2023knf,Lozanov:2023rcd,Lozanov:2022yoy}.
Hence, GW background spectrum can be much richer than previously thought.

\begin{figure}[t]
	\includegraphics[width=0.45\textwidth]{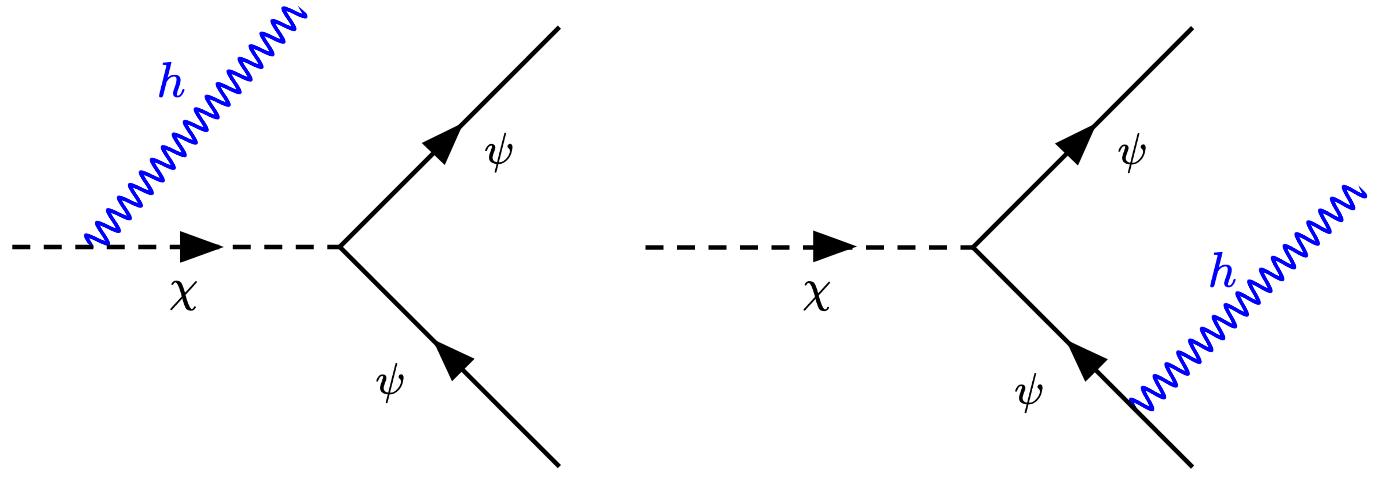}  
\caption{Example graviton bremsstrahlung decays $\chi\to\psi\bar\psi+h$ of a scalar $\chi$ (dashed) into fermions $\psi$ (solid), involving one graviton $h$
(wiggle, blue) in the final states.}
  \label{fig:diagram}
\end{figure}

Since particle decays are always accompanied by graviton emission, GWs associated with graviton bremsstrahlung are of particular interest among quantum gravitational processes~\cite{Nakayama:2018ptw,Huang:2019lgd,Barman:2023ymn,Bernal:2023wus}. However, such processes are expected to be highly suppressed. This can be readily seen by considering
a scalar particle $\chi$ decaying into light particles $\psi$ through the two-body decay $\chi\to\psi\bar\psi$ in the early Universe. This is necessarily accompanied by the graviton bremsstrahlung emission $\chi\to\psi\bar\psi+h$ as shown on Fig.~\ref{fig:diagram}, where $h$ denotes the graviton.
Then, the resulting GW abundance in the present day Universe is directly proportional to the branching ratio, given by~\cite{Nakayama:2018ptw,Huang:2019lgd,Barman:2023ymn,Bernal:2023wus}
\begin{align} \label{eq:branch}
	{\rm Br}_{\chi\to\psi\bar\psi h} \sim \mathcal O(10^{-2}) \times \left(\frac{m_\chi}{M_{\rm Pl}}\right)^2~,
\end{align}
where $m_\chi$ is the mass of $\chi$ and $M_{\rm Pl} = 2.4 \times 10^{18}$~GeV is the reduced Planck scale. Since this branching ratio is independent of any other coupling constants, significant GW production is expected only for large masses $m_\chi$. 
If $\chi$ is the inflaton particle that drives rapid early Universe expansion, to reproduce the density perturbations of the Universe most known inflation models predict the inflaton as heavy as few~$\times 10^{13}\,{\rm GeV}$~\cite{Liddle:2000cg}, implying branching ratio ${\rm Br}_{\chi\to\psi\bar\psi h} \lesssim 10^{-11}$. 

In this work we demonstrate that GW production and signatures from quantum gravitational bremsstrahlung processes can be dramatically enhanced over many orders of magnitude in frequency range. As we show, superheavy particles,
as heavy as the Planck scale, can be naturally produced during the post-inflationary reheating stage and copiously source GWs through graviton bremsstrahlung without suppression. Associated GWs can serve as direct probe of Planck-scale physics after inflation in the early Universe.
\\

\paragraph{Instant preheating.}--
After the period of inflation, driven by the inflaton, inflaton's energy density is converted into conventional matter and thus enters the Big Bang cosmology through reheating process. However, during the early stage of reheating, known as preheating, particles much heavier than the inflaton can be efficiently produced~\cite{Kofman:1994rk,Kofman:1997yn}. As we discuss, even temporary production of such superheavy particles and their subsequent decays can naturally result in significantly enhanced emission of GWs from graviton bremsstrahlung.

If a scalar particle $\chi$ is coupled to the inflaton, such particles are produced non-perturbatively through the inflaton oscillations after period of inflation. Since $\chi$'s mass depends on the field value of the inflaton, it is time-dependent. Then, $\chi$'s mass can be as heavy as the Planck scale temporarily.
If $\chi$ is further coupled to some lighter particle species, $\chi$ decays into them~\cite{Felder:1998vq}. This scenario of instant preheating allows for $\chi$ particles to be very efficiently produced while they decay when their masses are close to the Planck scale. This implies that branching ratio for processes in Eq.~\eqref{eq:branch} can naturally be as large as $\sim\mathcal O(0.1)$. 
Not only is this an ideal setting for efficient graviton bremsstrahlung but it also establishes a testing ground for Planck-scale physics, since the graviton energy may be close to the Planck scale.

This scenario can be readily realized within a minimal setup as we now show. Consider Lagrangian 
\begin{align}
	\mathcal L \supset -\frac{1}{2}m_\phi^2\phi^2 - \frac{1}{2}\lambda^2\phi^2\chi^2 + y\chi \bar\psi\psi~,
\end{align}
where $\phi$ is the inflaton, $\chi$ is a scalar field that is directly coupled to the inflaton through the four-point coupling $\lambda$, and $\psi$ is a fermion that has a Yukawa coupling to the $\chi$ particle with a coupling constant $y$. We have omitted canonical kinetic terms.

After inflation ends, the inflaton oscillates with an amplitude $\phi_i \sim M_{\rm Pl}$ and $\chi$ particles are non-perturbatively produced. The number density of the produced $\chi$ particles after just half of oscillation can be estimated as~\cite{Kofman:1997yn}
\begin{align}
	n_\chi \simeq \left(\frac{k_*}{2\pi}\right)^3~~,~~~~~~k_* = \sqrt{\lambda m_\phi \phi_i}~,
\end{align}
when $\lambda \phi_i \gg m_\phi$, with $k_{\ast}$ being the typical momenta. Produced $\chi$ particles can efficiently decay before the next half oscillation of the inflaton starts if the Yukawa coupling $y$ is sizable. In this scenario of instant preheating~\cite{Felder:1998vq}, sizable fraction of the inflaton energy rapidly transforms into radiation during just the first half of oscillation. 

From Yukawa coupling, the decay rate of $\chi$ is given by
\begin{align} 	\label{Gamma}
	\Gamma_{\chi\to\bar\psi\psi} = \frac{y^2 m_\chi}{8\pi} = \frac{y^2 \lambda \phi(t)}{8\pi}\simeq  \frac{y^2 \lambda m_\phi \phi_i t}{8\pi}~. 
\end{align}
Thus, the decay time is given by $t_{\rm dec} \simeq  \sqrt{8\pi/(y^2 \lambda m_\phi \phi_i)}$.
For $\chi$ particles to decay within half of oscillation, the decay time
should be smaller than the inflaton oscillation period $1/m_\phi$. Hence, we require
\begin{align} \label{y_constraint}
	m_\phi \lesssim \frac{ y^2 \lambda \phi_i}{8\pi}~,   
\end{align}
for the instant preheating to occur. Correspondingly, the mass of $\chi$ at the instance of its decay is given by
\begin{align} 	\label{mchi_dec}
	m_\chi (t_{\rm dec}) \simeq \lambda \phi(t_{\rm dec}) \simeq \sqrt{\frac{8\pi \lambda m_\phi \phi_i}{y^2}} \lesssim \lambda \phi_i~. 
\end{align}
When the upper bound of Eq.~\eqref{y_constraint} is saturated, the inflaton energy loss is most efficient. The energy loss of the inflaton in one half oscillation is then given by
\begin{align} 	\label{rhochi_rhophi}
	\frac{\delta\rho_\phi}{\rho_\phi} \sim \frac{m_\chi(t_{\rm dec}) n_\chi}{\rho_\phi}
	\sim \frac{\lambda^2}{4\pi^3}\sqrt{\frac{8\pi}{y^2}} \lesssim \mathcal O(1)\times \lambda^{5/2}~.
\end{align}
In the last inequality we used Eq.~(\ref{y_constraint}) and $m_\phi \sim 10^{13}\,{\rm GeV}$, $\phi_i \sim M_{\rm Pl}$. 

We note that effectiveness of our scenario depends on the strength of the coupling $\lambda$ between inflaton $\phi$ and $\chi$. This coupling also induces radiative correction to the inflaton potential through the Coleman-Weinberg potential $V_{\rm CW}\sim (32\pi^2)^{-1}(\lambda\phi)^4\log\phi$ .
If one is to require that inflaton dynamics are not spoiled by this correction in the setup at face value, this implies an upper bound $\lambda \lesssim 10^{-3}$.
Such restriction would result in inefficient particle production and also the effective mass of $\chi$ is bounded as $m_\chi \lesssim \lambda \phi_i \sim 10^{15}\,{\rm GeV}$. Hence, the graviton bremsstrahlung efficiency would also be suppressed.

We emphasize, however, that in many realistic inflation models such a constraint can be avoided and large enough $\lambda \sim \mathcal O(1)$ is achieved.
One such scenario is the Higgs inflation model and/or its variants~\cite{Bezrukov:2007ep}, in which the inflaton non-minimal coupling to the Ricci curvature ensures the flatness of the potential.
Another scenario is a setup based on supersymmetry (SUSY), in which the radiative correction is mostly cancelled out between the bosonic and fermionic contributions. 
For SUSY version of the instant preheating, we consider the Kahler potential $K$ and superpotential $W$ as
\begin{align}  
	& K = \frac{1}{2}(\phi+\phi^\dagger)^2 + |X|^2 + |\chi|^2 + |\psi|^2~, \label{K}\\
	& W = m_\phi X \phi + \lambda \phi \chi^2 + y \chi \psi^2~,   \label{W}
\end{align}
where $\phi$ is an inflaton superfield, $X$ is a stabilizer superfield, $\chi$ and $\psi$ are also superfields, containing both fermionic and bosonic components.
The inflaton superfield satisfies the approximate shift symmetry $\phi\to \phi + iC$ with arbitrary real constant $C$ so that the large field inflation works successfully in the supergravity framework~\cite{Kawasaki:2000yn}. We note in passing 
that to make predictions on the cosmological scalar spectral index and tensor-to-scalar ratio consistent with Planck  observations~\cite{Planck:2018jri}, modifications on the Kahler and/or superpotential need to be considered~\cite{Nakayama:2013jka,Nakayama:2013txa,Kallosh:2013hoa,Kallosh:2013yoa,Galante:2014ifa,Nakayama:2016gvg}.
However, such modifications only change the inflaton potential during inflation and have little effect on the reheating dynamics we are interested in here. Thus, we neglect this in the following. 

Eq.~(\ref{mchi_dec}) indicates that $\chi$ mass can be as heavy as $m_\chi \sim M_{\rm Pl}$ at the time of its decay.
In such a case, the significant graviton bremsstrahlung contributions associated with $\chi$ decays result in a large amount of GWs at high frequency range, as we show below.
Note also that Eq.~(\ref{rhochi_rhophi}) indicates that it is possible that the inflaton loses almost all of its energy within few oscillations and the radiation-dominated Universe begins soon after inflation.

We highlight that instant preheating readily occurs within multiple realistic models of inflation.
One example is the Higgs inflation~\cite{Bezrukov:2007ep}, where the gauge bosons take role of the $\chi$ particles that decay into quarks and leptons~\cite{Bezrukov:2008ut,Garcia-Bellido:2008ycs}. While reheating in the original Higgs inflation suffers from a unitarity issue~\cite{Ema:2016dny,DeCross:2016cbs,Sfakianakis:2018lzf,Ema:2021xhq}, this can be avoided in the Palatini Higgs inflation~\cite{Bauer:2008zj,Bauer:2010jg,Rubio:2019ypq}.
Furthermore, the SUSY model described by Eq.~(\ref{K}) and (\ref{W}) is naturally realized in the sneutrino chaotic inflation model~\cite{Murayama:2014saa,Evans:2015mta,Nakayama:2016gvg,Kallosh:2016sej}.
We regard the right-handed sneutrinos ($N_1,N_2$) as the inflaton or stabilizer, which couples to the up-type Higgs $H_u$ and lepton doublets $L_\alpha$ ($\alpha=1,2,3$) so that it explains the neutrino masses through the seesaw mechanism~\cite{Minkowski:1977sc,Yanagida:1979as,Gell-Mann:1979vob}. Here, $\chi$ is a combination of $L_\alpha$ and $H_u$, and $\psi$ collectively denote quarks, leptons or down-type Higgs. In this model instant preheating happens after inflation~\cite{Nakayama:2016gvg}.
Moreover, due to the unique structure of the right-handed neutrino mass matrix, such scenario can carry intriguing implications for neutrino charge-parity phases~\cite{Nakayama:2017cij}.
Thus, our setup can also have intimate connection with neutrino physics and leptogenesis~\cite{Fukugita:1986hr}.
As these examples indicate, if the inflaton is closely linked to the Standard Model fields, instant preheating can occur and hence significant GW production can be expected as we discuss. 
\\

\paragraph{Gravitational waves.}--
We now compute the resulting GW spectrum from our scenario, sourced by graviton bremsstrahlung.
During the first half of inflaton 
 oscillation just after inflation, $\chi$ particles are efficiently produced and decay into $\psi$ pairs with the rate given by Eq.~\eqref{Gamma}.
This two-body decay is associated with the three-body graviton bremsstrahlung process $\chi\to \psi \bar\psi + h$, whose differential decay rate over $h$'s energy $E$  is given by~\cite{Nakayama:2018ptw,Huang:2019lgd,Barman:2023ymn}
\begin{align} 	\label{dGdE}
	\frac{d\Gamma_{\rm brem}}{d\ln E} =\frac{y^2}{64\pi^3}\frac{m_\chi^3}{M_{\rm Pl}^2}(1-2x)(1-2x+2x^2)~,
\end{align}
where $x\equiv E/m_\chi$. Thus, one can see that the branching ratio of $\chi$ to the graviton bremsstrahlung is
\begin{align}
	{\rm Br}_{\chi\to\psi\bar\psi h}\simeq \frac{1}{8\pi^2} \left(\frac{m_\chi}{M_{\rm Pl}}\right)^2 
\end{align}
and it depends only on the mass of $\chi$. Therefore, heavier $\chi$ will result in more significant amount of produced GWs in the Universe.
As we have shown above, $m_\chi$ can naturally be as heavy as $M_{\rm Pl}$ during preheating and hence we can expect very large GW signals in this case.

The present day energy density spectrum associated with bremsstrahlung graviton emission, with $E_0$ being the present energy, is given by
\begin{align}
	\frac{d\rho_{\rm GW}}{d\ln E_0}
	= E_0\int dt \, n_\chi(t) \left(\frac{a(t)}{a_0}\right)^3 \frac{d\Gamma_{\rm brem}(t)}{d\ln E}~,
\end{align}
where $\rho_{\rm GW}$ is the GW energy density, $n_\chi$ is the number density of $\chi$ as before and $E(t)=E_0 a_0/a(t)$, with $a(t)$ being the cosmological scale factor and the subscript 0 indicating its present value.
Here, we consider instantaneous decay approximation, in which $\chi$ completely decays within one Hubble time, i.e. $\Gamma_{\chi\to\psi\bar\psi} \gg H$ at the corresponding redshift $z=z_{\rm dec}$ with $H$ being the Hubble expansion rate. 

Then, the present-day GW spectrum $d\Omega_{\rm GW}/d\ln E_0 = (1/\rho_c) d \rho_{\rm GW}/ d\ln E_0$, where $\rho_c =  1.05 \times 10^{-5} h^2$~GeV/cm$^{3}$
is the critical energy density at present and $h \simeq 0.67$ is the present dimensionless Hubble parameter,
can be found to be
\begin{align}	\label{Ogw}
	\frac{d\Omega_{\rm GW}}{d\ln E_0} 
	=&~\epsilon\,\Omega_{\rm rad} \left(\frac{g_*}{g_{*0}}\right)\left(\frac{g_{*s0}}{g_{*s}}\right)^{\frac{4}{3}} 
	\left(\frac{m_\chi n_\chi}{\rho_{\rm tot}}\right)_{z_{\rm dec}} \\ 
	&\times {\rm Br}_{\chi\to\psi\bar\psi h}
	\times x (1-2x)(1-2x+2x^2)~, \notag
\end{align}
where 
$x= E_0/E_p$ with $E_p \equiv m_\chi/(1+z_{\rm dec})$, $\rho_{\rm tot}$ denotes the total Universe energy density, $\Omega_{\rm rad}\sim 8\times 10^{-5}$ is the present-day radiation density parameter, $g_*$ and $g_{*s}$ are the effective relativistic degrees of freedom for the energy and entropy density respectively.
The factor $(m_\chi n_\chi/\rho_{\rm tot})_{z_{\rm dec}}$ is given by Eq.~\eqref{rhochi_rhophi}.
The overall prefactor $\epsilon$ takes into account the equation of state of the Universe $w$ until reheating is completed. It is given by $\epsilon = \left(a_{\rm dec}/a_{\rm reh}\right)^{1-3w}$,
where $a_{\rm dec}$ and $a_{\rm reh}$ denote the scale factor at the $\chi$ decay and the completion of reheating, respectively.
In the limit of most efficient instant preheating, the Universe soon turns into radiation domination after inflation, so $w=1/3$ and $\epsilon=1$. Inefficient prolonged reheating phase would result in small $\epsilon$. 
The peak value of the GW spectrum in Eq.~\eqref{Ogw} at energy $E_p$ approximately follows $\Omega_{\rm GW} \sim \epsilon\Omega_{\rm rad} {\rm Br}_{\chi\to\psi\bar\psi h}  (\rho_\chi/\rho_{\rm tot})_{z_{\rm dec}}$. For $E_0 \lesssim E_p$, the spectrum is proportional to $E_0$.
Here, we do not go into details regarding the exact range how far the spectrum of Eq.~\eqref{Ogw} can be extended to low frequencies~\cite{Nakayama:2018ptw}.
When the upper bound on Eq.~\eqref{mchi_dec} is saturated, we have $ (\rho_\chi/\rho_{\rm tot})_{z_{\rm dec}} \sim \mathcal O(1)$ and $\epsilon=1$, and hence we can expect huge generated amount of GWs.
The typical GW frequency, $f_p=E_p/(2\pi)$, can be estimated as
\begin{align}
	f_p \simeq 4.6\times 10^{13}\,{\rm Hz}\times \left(\frac{a_{\rm dec}}{a_{\rm reh}}\right)\left(\frac{m_\chi}{M_{\rm Pl}}\right) \left(\frac{10^{15}\,{\rm GeV}}{T_{\rm reh}}\right)~,
\end{align}
where $T_{\rm reh}$ is the reheating temperature. 

\begin{figure}[t]
	\includegraphics[width=0.475 \textwidth]{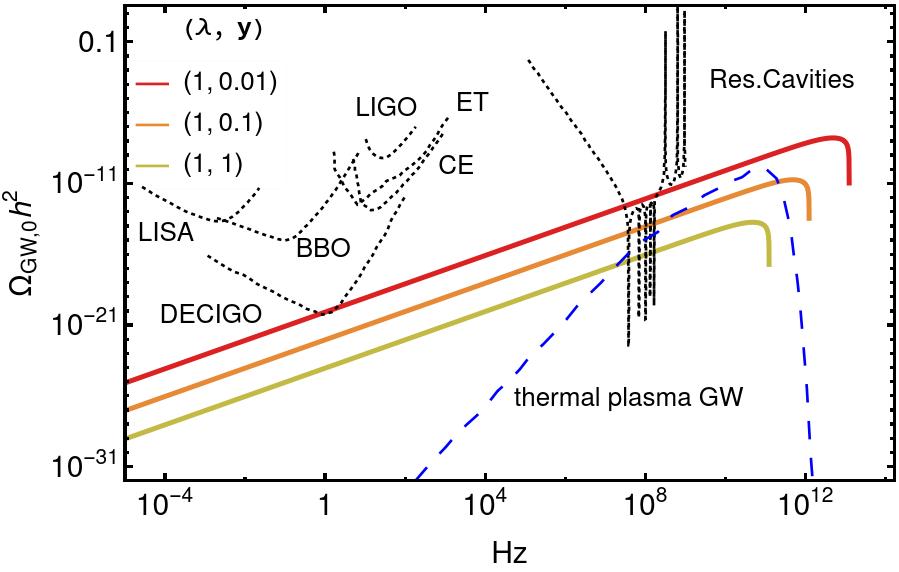}  
\caption{Expected GW energy spectrum at present time from graviton bremsstrahlung associated with heavy particle decays during reheating for several choices of  $(\lambda, y)$ (solid), along with sensitivity curves (black, dashed) of proposed and existing GW detectors, the fifth observing run (O5) of the aLIGO-Virgo detector network~\cite{LIGOScientific:2016fpe}, Laser Interferometer Space Antenna (LISA)~\cite{LISA:2017pwj}, Cosmic Explorer (CE)~\cite{Reitze:2019iox}, Einstein Telescope (ET)~\cite{Punturo:2010zz}, Big Bang Observer (BBO)~\cite{Harry:2006fi}, DECi-hertz Interferometer Gravitational wave Observatory (DECIGO)~\cite{Seto:2001qf} and resonance cavities~\cite{Herman:2020wao,Herman:2022fau}. Contributions from thermal plasma~\cite{Ghiglieri:2015nfa,Ghiglieri:2020mhm,Ringwald:2020ist} (blue, dashed) are displayed for reference.}
  \label{fig:GWdecay}
\end{figure}

Fig.~\ref{fig:GWdecay} shows GW energy spectra from graviton bremsstrahlung in our scenario for several characteristic parameter choices of $(\lambda, y)$ along with sensitivities of proposed and existing GW detectors, considering $m_\phi=10^{13}\,{\rm GeV}$ and $\phi_i=M_{\rm Pl}$.
Here, we consider efficient preheating such that the radiation-dominated Universe begins soon after inflation ends, with $\lambda y^2\gtrsim 10^{-4}$ in Eq.~\eqref{y_constraint}.
For reference, we also showcase the expected GW contributions from thermal plasma~\cite{Ghiglieri:2015nfa,Ghiglieri:2020mhm,Ringwald:2020ist} taking the maximum reheating temperature of $T_{\rm reh}=10^{16}\,{\rm GeV}$.
It can be seen that the peak position of the GW energy spectrum is determined by $m_{\chi}$ and peak frequency ranges $\sim10^9 - 10^{13}$\,Hz. Over broad range in frequencies, bremsstrahlung GWs can dominate thermal plasma GWs by orders of magnitude.
Intriguingly, recently proposed GW detectors based on resonant cavities~\cite{Herman:2020wao,Herman:2022fau} are sensitive to GWs from graviton bremsstrahlung due to heavy particle decays during instant preheating. Further, at lower frequencies, our scenario can also be tested by proposed GW experiments such as DECIGO~\cite{Seto:2001qf}.

We have demonstrated multiple venues for successful realizations of instant preheating. On the other hand, scenarios with small couplings between inflaton and $\chi$, when the inequality of Eq.~\eqref{y_constraint} is not satisfied, are also interesting to further explore. 
In such a case, most of $\chi$ particles do not decay in one inflaton oscillation period and resulting Bose enhancement can lead to broad parametric resonance~\cite{Kofman:1994rk,Kofman:1997yn}.
Then we expect mixture of GWs from the bremsstrahlung and that from the large density fluctuation of the scalar field~\cite{Khlebnikov:1997di,Easther:2006gt,Easther:2006vd,Garcia-Bellido:2007fiu,Dufaux:2007pt,Figueroa:2011ye,Repond:2016sol,Figueroa:2017vfa,Figueroa:2022iho}.
We leave detailed investigation of this for future work.  
\\

\paragraph{Conclusions.}--
We have demonstrated that quantum gravitational processes accompanying all particle decays can be unsuppressed and produce huge amounts of GWs. As we have shown with concrete particle physics models, this can naturally occur for graviton bremsstrahlung from superheavy particle decays within preheating stage after period of inflation in the early Universe. The masses of such decaying particles can be close to the Planck scale in our scenario and hence the observation of resulting GWs can be a direct probe of the Planck-scale physics that are challenging to probe otherwise. 
Interestingly, such GWs can dominate among other known GW sources over a wide frequency range and are within the reach of recently proposed experiments.
On the other hand, if there are significant yet unexplored modifications of quantum field theory close to the Planck scale, then the decay rates, spectra or the graviton dispersion relations could be altered and affect resulting GW spectra. 
Thus, observations of high frequency GWs can carry unique opportunities for new insights into the Planck-scale physics.
\\

\paragraph{Acknowledgments.}--
This work was supported by World Premier International Research Center Initiative (WPI), MEXT, Japan. W.Y.H was supported by the National Natural Science Foundation of China (NSFC) (No. 11975029, No. 12325503). V.T. acknowledges support of the JSPS KAKENHI grant No. 23K13109. Y.T. was supported by NSFC No.~12147103 and the Fundamental Research Funds for the Central Universities.

\bibliographystyle{apsrev4-1}
\bibliography{ref}

\begin{thebibliography}{80}%
\makeatletter
\providecommand \@ifxundefined [1]{%
 \@ifx{#1\undefined}
}%
\providecommand \@ifnum [1]{%
 \ifnum #1\expandafter \@firstoftwo
 \else \expandafter \@secondoftwo
 \fi
}%
\providecommand \@ifx [1]{%
 \ifx #1\expandafter \@firstoftwo
 \else \expandafter \@secondoftwo
 \fi
}%
\providecommand \natexlab [1]{#1}%
\providecommand \enquote  [1]{``#1''}%
\providecommand \bibnamefont  [1]{#1}%
\providecommand \bibfnamefont [1]{#1}%
\providecommand \citenamefont [1]{#1}%
\providecommand \href@noop [0]{\@secondoftwo}%
\providecommand \href [0]{\begingroup \@sanitize@url \@href}%
\providecommand \@href[1]{\@@startlink{#1}\@@href}%
\providecommand \@@href[1]{\endgroup#1\@@endlink}%
\providecommand \@sanitize@url [0]{\catcode `\\12\catcode `\$12\catcode `\&12\catcode `\#12\catcode `\^12\catcode `\_12\catcode `\%12\relax}%
\providecommand \@@startlink[1]{}%
\providecommand \@@endlink[0]{}%
\providecommand \url  [0]{\begingroup\@sanitize@url \@url }%
\providecommand \@url [1]{\endgroup\@href {#1}{\urlprefix }}%
\providecommand \urlprefix  [0]{URL }%
\providecommand \Eprint [0]{\href }%
\providecommand \doibase [0]{http://dx.doi.org/}%
\providecommand \selectlanguage [0]{\@gobble}%
\providecommand \bibinfo  [0]{\@secondoftwo}%
\providecommand \bibfield  [0]{\@secondoftwo}%
\providecommand \translation [1]{[#1]}%
\providecommand \BibitemOpen [0]{}%
\providecommand \bibitemStop [0]{}%
\providecommand \bibitemNoStop [0]{.\EOS\space}%
\providecommand \EOS [0]{\spacefactor3000\relax}%
\providecommand \BibitemShut  [1]{\csname bibitem#1\endcsname}%
\let\auto@bib@innerbib\@empty
\bibitem [{\citenamefont {Abbott}\ \emph {et~al.}(2016{\natexlab{a}})\citenamefont {Abbott} \emph {et~al.}}]{LIGOScientific:2016aoc}%
  \BibitemOpen
  \bibfield  {author} {\bibinfo {author} {\bibfnamefont {B.~P.}\ \bibnamefont {Abbott}} \emph {et~al.} (\bibinfo {collaboration} {LIGO Scientific, Virgo}),\ }\href {\doibase 10.1103/PhysRevLett.116.061102} {\bibfield  {journal} {\bibinfo  {journal} {Phys. Rev. Lett.}\ }\textbf {\bibinfo {volume} {116}},\ \bibinfo {pages} {061102} (\bibinfo {year} {2016}{\natexlab{a}})},\ \Eprint {http://arxiv.org/abs/1602.03837} {arXiv:1602.03837 [gr-qc]} \BibitemShut {NoStop}%
\bibitem [{\citenamefont {Abbott}\ \emph {et~al.}(2017)\citenamefont {Abbott} \emph {et~al.}}]{LIGOScientific:2017vwq}%
  \BibitemOpen
  \bibfield  {author} {\bibinfo {author} {\bibfnamefont {B.~P.}\ \bibnamefont {Abbott}} \emph {et~al.} (\bibinfo {collaboration} {LIGO Scientific, Virgo}),\ }\href {\doibase 10.1103/PhysRevLett.119.161101} {\bibfield  {journal} {\bibinfo  {journal} {Phys. Rev. Lett.}\ }\textbf {\bibinfo {volume} {119}},\ \bibinfo {pages} {161101} (\bibinfo {year} {2017})},\ \Eprint {http://arxiv.org/abs/1710.05832} {arXiv:1710.05832 [gr-qc]} \BibitemShut {NoStop}%
\bibitem [{\citenamefont {Agazie}\ \emph {et~al.}(2023)\citenamefont {Agazie} \emph {et~al.}}]{NANOGrav:2023gor}%
  \BibitemOpen
  \bibfield  {author} {\bibinfo {author} {\bibfnamefont {G.}~\bibnamefont {Agazie}} \emph {et~al.} (\bibinfo {collaboration} {NANOGrav}),\ }\href {\doibase 10.3847/2041-8213/acdac6} {\bibfield  {journal} {\bibinfo  {journal} {Astrophys. J. Lett.}\ }\textbf {\bibinfo {volume} {951}},\ \bibinfo {pages} {L8} (\bibinfo {year} {2023})},\ \Eprint {http://arxiv.org/abs/2306.16213} {arXiv:2306.16213 [astro-ph.HE]} \BibitemShut {NoStop}%
\bibitem [{\citenamefont {Xu}\ \emph {et~al.}(2023)\citenamefont {Xu} \emph {et~al.}}]{Xu:2023wog}%
  \BibitemOpen
  \bibfield  {author} {\bibinfo {author} {\bibfnamefont {H.}~\bibnamefont {Xu}} \emph {et~al.},\ }\href {\doibase 10.1088/1674-4527/acdfa5} {\bibfield  {journal} {\bibinfo  {journal} {Res. Astron. Astrophys.}\ }\textbf {\bibinfo {volume} {23}},\ \bibinfo {pages} {075024} (\bibinfo {year} {2023})},\ \Eprint {http://arxiv.org/abs/2306.16216} {arXiv:2306.16216 [astro-ph.HE]} \BibitemShut {NoStop}%
\bibitem [{\citenamefont {Reardon}\ \emph {et~al.}(2023)\citenamefont {Reardon} \emph {et~al.}}]{Reardon:2023gzh}%
  \BibitemOpen
  \bibfield  {author} {\bibinfo {author} {\bibfnamefont {D.~J.}\ \bibnamefont {Reardon}} \emph {et~al.},\ }\href {\doibase 10.3847/2041-8213/acdd02} {\bibfield  {journal} {\bibinfo  {journal} {Astrophys. J. Lett.}\ }\textbf {\bibinfo {volume} {951}},\ \bibinfo {pages} {L6} (\bibinfo {year} {2023})},\ \Eprint {http://arxiv.org/abs/2306.16215} {arXiv:2306.16215 [astro-ph.HE]} \BibitemShut {NoStop}%
\bibitem [{\citenamefont {Antoniadis}\ \emph {et~al.}(2023)\citenamefont {Antoniadis} \emph {et~al.}}]{EPTA:2023fyk}%
  \BibitemOpen
  \bibfield  {author} {\bibinfo {author} {\bibfnamefont {J.}~\bibnamefont {Antoniadis}} \emph {et~al.} (\bibinfo {collaboration} {EPTA, InPTA:}),\ }\href {\doibase 10.1051/0004-6361/202346844} {\bibfield  {journal} {\bibinfo  {journal} {Astron. Astrophys.}\ }\textbf {\bibinfo {volume} {678}},\ \bibinfo {pages} {A50} (\bibinfo {year} {2023})},\ \Eprint {http://arxiv.org/abs/2306.16214} {arXiv:2306.16214 [astro-ph.HE]} \BibitemShut {NoStop}%
\bibitem [{\citenamefont {Maggiore}(2018)}]{Maggiore:2018sht}%
  \BibitemOpen
  \bibfield  {author} {\bibinfo {author} {\bibfnamefont {M.}~\bibnamefont {Maggiore}},\ }\href@noop {} {\emph {\bibinfo {title} {{Gravitational Waves. Vol. 2: Astrophysics and Cosmology}}}}\ (\bibinfo  {publisher} {Oxford University Press},\ \bibinfo {year} {2018})\BibitemShut {NoStop}%
\bibitem [{\citenamefont {Ghiglieri}\ and\ \citenamefont {Laine}(2015)}]{Ghiglieri:2015nfa}%
  \BibitemOpen
  \bibfield  {author} {\bibinfo {author} {\bibfnamefont {J.}~\bibnamefont {Ghiglieri}}\ and\ \bibinfo {author} {\bibfnamefont {M.}~\bibnamefont {Laine}},\ }\href {\doibase 10.1088/1475-7516/2015/07/022} {\bibfield  {journal} {\bibinfo  {journal} {JCAP}\ }\textbf {\bibinfo {volume} {07}},\ \bibinfo {pages} {022} (\bibinfo {year} {2015})},\ \Eprint {http://arxiv.org/abs/1504.02569} {arXiv:1504.02569 [hep-ph]} \BibitemShut {NoStop}%
\bibitem [{\citenamefont {Ghiglieri}\ \emph {et~al.}(2020)\citenamefont {Ghiglieri}, \citenamefont {Jackson}, \citenamefont {Laine},\ and\ \citenamefont {Zhu}}]{Ghiglieri:2020mhm}%
  \BibitemOpen
  \bibfield  {author} {\bibinfo {author} {\bibfnamefont {J.}~\bibnamefont {Ghiglieri}}, \bibinfo {author} {\bibfnamefont {G.}~\bibnamefont {Jackson}}, \bibinfo {author} {\bibfnamefont {M.}~\bibnamefont {Laine}}, \ and\ \bibinfo {author} {\bibfnamefont {Y.}~\bibnamefont {Zhu}},\ }\href {\doibase 10.1007/JHEP07(2020)092} {\bibfield  {journal} {\bibinfo  {journal} {JHEP}\ }\textbf {\bibinfo {volume} {07}},\ \bibinfo {pages} {092} (\bibinfo {year} {2020})},\ \Eprint {http://arxiv.org/abs/2004.11392} {arXiv:2004.11392 [hep-ph]} \BibitemShut {NoStop}%
\bibitem [{\citenamefont {Ringwald}\ \emph {et~al.}(2021)\citenamefont {Ringwald}, \citenamefont {Sch\"utte-Engel},\ and\ \citenamefont {Tamarit}}]{Ringwald:2020ist}%
  \BibitemOpen
  \bibfield  {author} {\bibinfo {author} {\bibfnamefont {A.}~\bibnamefont {Ringwald}}, \bibinfo {author} {\bibfnamefont {J.}~\bibnamefont {Sch\"utte-Engel}}, \ and\ \bibinfo {author} {\bibfnamefont {C.}~\bibnamefont {Tamarit}},\ }\href {\doibase 10.1088/1475-7516/2021/03/054} {\bibfield  {journal} {\bibinfo  {journal} {JCAP}\ }\textbf {\bibinfo {volume} {03}},\ \bibinfo {pages} {054} (\bibinfo {year} {2021})},\ \Eprint {http://arxiv.org/abs/2011.04731} {arXiv:2011.04731 [hep-ph]} \BibitemShut {NoStop}%
\bibitem [{\citenamefont {Nakayama}\ and\ \citenamefont {Tang}(2019)}]{Nakayama:2018ptw}%
  \BibitemOpen
  \bibfield  {author} {\bibinfo {author} {\bibfnamefont {K.}~\bibnamefont {Nakayama}}\ and\ \bibinfo {author} {\bibfnamefont {Y.}~\bibnamefont {Tang}},\ }\href {\doibase 10.1016/j.physletb.2018.11.023} {\bibfield  {journal} {\bibinfo  {journal} {Phys. Lett. B}\ }\textbf {\bibinfo {volume} {788}},\ \bibinfo {pages} {341} (\bibinfo {year} {2019})},\ \Eprint {http://arxiv.org/abs/1810.04975} {arXiv:1810.04975 [hep-ph]} \BibitemShut {NoStop}%
\bibitem [{\citenamefont {Huang}\ and\ \citenamefont {Yin}(2019)}]{Huang:2019lgd}%
  \BibitemOpen
  \bibfield  {author} {\bibinfo {author} {\bibfnamefont {D.}~\bibnamefont {Huang}}\ and\ \bibinfo {author} {\bibfnamefont {L.}~\bibnamefont {Yin}},\ }\href {\doibase 10.1103/PhysRevD.100.043538} {\bibfield  {journal} {\bibinfo  {journal} {Phys. Rev. D}\ }\textbf {\bibinfo {volume} {100}},\ \bibinfo {pages} {043538} (\bibinfo {year} {2019})},\ \Eprint {http://arxiv.org/abs/1905.08510} {arXiv:1905.08510 [hep-ph]} \BibitemShut {NoStop}%
\bibitem [{\citenamefont {Barman}\ \emph {et~al.}(2023)\citenamefont {Barman}, \citenamefont {Bernal}, \citenamefont {Xu},\ and\ \citenamefont {Zapata}}]{Barman:2023ymn}%
  \BibitemOpen
  \bibfield  {author} {\bibinfo {author} {\bibfnamefont {B.}~\bibnamefont {Barman}}, \bibinfo {author} {\bibfnamefont {N.}~\bibnamefont {Bernal}}, \bibinfo {author} {\bibfnamefont {Y.}~\bibnamefont {Xu}}, \ and\ \bibinfo {author} {\bibfnamefont {O.}~\bibnamefont {Zapata}},\ }\href {\doibase 10.1088/1475-7516/2023/05/019} {\bibfield  {journal} {\bibinfo  {journal} {JCAP}\ }\textbf {\bibinfo {volume} {05}},\ \bibinfo {pages} {019} (\bibinfo {year} {2023})},\ \Eprint {http://arxiv.org/abs/2301.11345} {arXiv:2301.11345 [hep-ph]} \BibitemShut {NoStop}%
\bibitem [{\citenamefont {Bernal}\ \emph {et~al.}(2024)\citenamefont {Bernal}, \citenamefont {Cl\'ery}, \citenamefont {Mambrini},\ and\ \citenamefont {Xu}}]{Bernal:2023wus}%
  \BibitemOpen
  \bibfield  {author} {\bibinfo {author} {\bibfnamefont {N.}~\bibnamefont {Bernal}}, \bibinfo {author} {\bibfnamefont {S.}~\bibnamefont {Cl\'ery}}, \bibinfo {author} {\bibfnamefont {Y.}~\bibnamefont {Mambrini}}, \ and\ \bibinfo {author} {\bibfnamefont {Y.}~\bibnamefont {Xu}},\ }\href {\doibase 10.1088/1475-7516/2024/01/065} {\bibfield  {journal} {\bibinfo  {journal} {JCAP}\ }\textbf {\bibinfo {volume} {01}},\ \bibinfo {pages} {065} (\bibinfo {year} {2024})},\ \Eprint {http://arxiv.org/abs/2311.12694} {arXiv:2311.12694 [hep-ph]} \BibitemShut {NoStop}%
\bibitem [{\citenamefont {Ema}\ \emph {et~al.}(2015)\citenamefont {Ema}, \citenamefont {Jinno}, \citenamefont {Mukaida},\ and\ \citenamefont {Nakayama}}]{Ema:2015dka}%
  \BibitemOpen
  \bibfield  {author} {\bibinfo {author} {\bibfnamefont {Y.}~\bibnamefont {Ema}}, \bibinfo {author} {\bibfnamefont {R.}~\bibnamefont {Jinno}}, \bibinfo {author} {\bibfnamefont {K.}~\bibnamefont {Mukaida}}, \ and\ \bibinfo {author} {\bibfnamefont {K.}~\bibnamefont {Nakayama}},\ }\href {\doibase 10.1088/1475-7516/2015/05/038} {\bibfield  {journal} {\bibinfo  {journal} {JCAP}\ }\textbf {\bibinfo {volume} {05}},\ \bibinfo {pages} {038} (\bibinfo {year} {2015})},\ \Eprint {http://arxiv.org/abs/1502.02475} {arXiv:1502.02475 [hep-ph]} \BibitemShut {NoStop}%
\bibitem [{\citenamefont {Ema}\ \emph {et~al.}(2016)\citenamefont {Ema}, \citenamefont {Jinno}, \citenamefont {Mukaida},\ and\ \citenamefont {Nakayama}}]{Ema:2016hlw}%
  \BibitemOpen
  \bibfield  {author} {\bibinfo {author} {\bibfnamefont {Y.}~\bibnamefont {Ema}}, \bibinfo {author} {\bibfnamefont {R.}~\bibnamefont {Jinno}}, \bibinfo {author} {\bibfnamefont {K.}~\bibnamefont {Mukaida}}, \ and\ \bibinfo {author} {\bibfnamefont {K.}~\bibnamefont {Nakayama}},\ }\href {\doibase 10.1103/PhysRevD.94.063517} {\bibfield  {journal} {\bibinfo  {journal} {Phys. Rev. D}\ }\textbf {\bibinfo {volume} {94}},\ \bibinfo {pages} {063517} (\bibinfo {year} {2016})},\ \Eprint {http://arxiv.org/abs/1604.08898} {arXiv:1604.08898 [hep-ph]} \BibitemShut {NoStop}%
\bibitem [{\citenamefont {Ema}\ \emph {et~al.}(2020)\citenamefont {Ema}, \citenamefont {Jinno},\ and\ \citenamefont {Nakayama}}]{Ema:2020ggo}%
  \BibitemOpen
  \bibfield  {author} {\bibinfo {author} {\bibfnamefont {Y.}~\bibnamefont {Ema}}, \bibinfo {author} {\bibfnamefont {R.}~\bibnamefont {Jinno}}, \ and\ \bibinfo {author} {\bibfnamefont {K.}~\bibnamefont {Nakayama}},\ }\href {\doibase 10.1088/1475-7516/2020/09/015} {\bibfield  {journal} {\bibinfo  {journal} {JCAP}\ }\textbf {\bibinfo {volume} {09}},\ \bibinfo {pages} {015} (\bibinfo {year} {2020})},\ \Eprint {http://arxiv.org/abs/2006.09972} {arXiv:2006.09972 [astro-ph.CO]} \BibitemShut {NoStop}%
\bibitem [{\citenamefont {Ghiglieri}\ \emph {et~al.}(2024{\natexlab{a}})\citenamefont {Ghiglieri}, \citenamefont {Sch\"utte-Engel},\ and\ \citenamefont {Speranza}}]{Ghiglieri:2022rfp}%
  \BibitemOpen
  \bibfield  {author} {\bibinfo {author} {\bibfnamefont {J.}~\bibnamefont {Ghiglieri}}, \bibinfo {author} {\bibfnamefont {J.}~\bibnamefont {Sch\"utte-Engel}}, \ and\ \bibinfo {author} {\bibfnamefont {E.}~\bibnamefont {Speranza}},\ }\href {\doibase 10.1103/PhysRevD.109.023538} {\bibfield  {journal} {\bibinfo  {journal} {Phys. Rev. D}\ }\textbf {\bibinfo {volume} {109}},\ \bibinfo {pages} {023538} (\bibinfo {year} {2024}{\natexlab{a}})},\ \Eprint {http://arxiv.org/abs/2211.16513} {arXiv:2211.16513 [hep-ph]} \BibitemShut {NoStop}%
\bibitem [{\citenamefont {Ghiglieri}\ \emph {et~al.}(2024{\natexlab{b}})\citenamefont {Ghiglieri}, \citenamefont {Laine}, \citenamefont {Sch\"utte-Engel},\ and\ \citenamefont {Speranza}}]{Ghiglieri:2024ghm}%
  \BibitemOpen
  \bibfield  {author} {\bibinfo {author} {\bibfnamefont {J.}~\bibnamefont {Ghiglieri}}, \bibinfo {author} {\bibfnamefont {M.}~\bibnamefont {Laine}}, \bibinfo {author} {\bibfnamefont {J.}~\bibnamefont {Sch\"utte-Engel}}, \ and\ \bibinfo {author} {\bibfnamefont {E.}~\bibnamefont {Speranza}},\ }\href@noop {} {\  (\bibinfo {year} {2024}{\natexlab{b}})},\ \Eprint {http://arxiv.org/abs/2401.08766} {arXiv:2401.08766 [hep-ph]} \BibitemShut {NoStop}%
\bibitem [{\citenamefont {Choi}\ \emph {et~al.}(2024)\citenamefont {Choi}, \citenamefont {Ke},\ and\ \citenamefont {Olive}}]{Choi:2024ilx}%
  \BibitemOpen
  \bibfield  {author} {\bibinfo {author} {\bibfnamefont {G.}~\bibnamefont {Choi}}, \bibinfo {author} {\bibfnamefont {W.}~\bibnamefont {Ke}}, \ and\ \bibinfo {author} {\bibfnamefont {K.~A.}\ \bibnamefont {Olive}},\ }\href@noop {} {\  (\bibinfo {year} {2024})},\ \Eprint {http://arxiv.org/abs/2402.04310} {arXiv:2402.04310 [hep-ph]} \BibitemShut {NoStop}%
\bibitem [{\citenamefont {Ema}\ \emph {et~al.}(2022)\citenamefont {Ema}, \citenamefont {Mukaida},\ and\ \citenamefont {Nakayama}}]{Ema:2021fdz}%
  \BibitemOpen
  \bibfield  {author} {\bibinfo {author} {\bibfnamefont {Y.}~\bibnamefont {Ema}}, \bibinfo {author} {\bibfnamefont {K.}~\bibnamefont {Mukaida}}, \ and\ \bibinfo {author} {\bibfnamefont {K.}~\bibnamefont {Nakayama}},\ }\href {\doibase 10.1007/JHEP05(2022)087} {\bibfield  {journal} {\bibinfo  {journal} {JHEP}\ }\textbf {\bibinfo {volume} {05}},\ \bibinfo {pages} {087} (\bibinfo {year} {2022})},\ \Eprint {http://arxiv.org/abs/2112.12774} {arXiv:2112.12774 [hep-ph]} \BibitemShut {NoStop}%
\bibitem [{\citenamefont {Mudrunka}\ and\ \citenamefont {Nakayama}(2023)}]{Mudrunka:2023wxy}%
  \BibitemOpen
  \bibfield  {author} {\bibinfo {author} {\bibfnamefont {K.}~\bibnamefont {Mudrunka}}\ and\ \bibinfo {author} {\bibfnamefont {K.}~\bibnamefont {Nakayama}},\ }\href@noop {} {\  (\bibinfo {year} {2023})},\ \Eprint {http://arxiv.org/abs/2312.15766} {arXiv:2312.15766 [astro-ph.CO]} \BibitemShut {NoStop}%
\bibitem [{\citenamefont {Tokareva}(2023)}]{Tokareva:2023mrt}%
  \BibitemOpen
  \bibfield  {author} {\bibinfo {author} {\bibfnamefont {A.}~\bibnamefont {Tokareva}},\ }\href@noop {} {\  (\bibinfo {year} {2023})},\ \Eprint {http://arxiv.org/abs/2312.16691} {arXiv:2312.16691 [hep-ph]} \BibitemShut {NoStop}%
\bibitem [{\citenamefont {Zhou}\ \emph {et~al.}(2013)\citenamefont {Zhou}, \citenamefont {Copeland}, \citenamefont {Easther}, \citenamefont {Finkel}, \citenamefont {Mou},\ and\ \citenamefont {Saffin}}]{Zhou:2013tsa}%
  \BibitemOpen
  \bibfield  {author} {\bibinfo {author} {\bibfnamefont {S.-Y.}\ \bibnamefont {Zhou}}, \bibinfo {author} {\bibfnamefont {E.~J.}\ \bibnamefont {Copeland}}, \bibinfo {author} {\bibfnamefont {R.}~\bibnamefont {Easther}}, \bibinfo {author} {\bibfnamefont {H.}~\bibnamefont {Finkel}}, \bibinfo {author} {\bibfnamefont {Z.-G.}\ \bibnamefont {Mou}}, \ and\ \bibinfo {author} {\bibfnamefont {P.~M.}\ \bibnamefont {Saffin}},\ }\href {\doibase 10.1007/JHEP10(2013)026} {\bibfield  {journal} {\bibinfo  {journal} {JHEP}\ }\textbf {\bibinfo {volume} {10}},\ \bibinfo {pages} {026} (\bibinfo {year} {2013})},\ \Eprint {http://arxiv.org/abs/1304.6094} {arXiv:1304.6094 [astro-ph.CO]} \BibitemShut {NoStop}%
\bibitem [{\citenamefont {Antusch}\ \emph {et~al.}(2017)\citenamefont {Antusch}, \citenamefont {Cefala},\ and\ \citenamefont {Orani}}]{Antusch:2016con}%
  \BibitemOpen
  \bibfield  {author} {\bibinfo {author} {\bibfnamefont {S.}~\bibnamefont {Antusch}}, \bibinfo {author} {\bibfnamefont {F.}~\bibnamefont {Cefala}}, \ and\ \bibinfo {author} {\bibfnamefont {S.}~\bibnamefont {Orani}},\ }\href {\doibase 10.1103/PhysRevLett.118.011303} {\bibfield  {journal} {\bibinfo  {journal} {Phys. Rev. Lett.}\ }\textbf {\bibinfo {volume} {118}},\ \bibinfo {pages} {011303} (\bibinfo {year} {2017})},\ \bibinfo {note} {[Erratum: Phys.Rev.Lett. 120, 219901 (2018)]},\ \Eprint {http://arxiv.org/abs/1607.01314} {arXiv:1607.01314 [astro-ph.CO]} \BibitemShut {NoStop}%
\bibitem [{\citenamefont {Liu}\ \emph {et~al.}(2018)\citenamefont {Liu}, \citenamefont {Guo}, \citenamefont {Cai},\ and\ \citenamefont {Shiu}}]{Liu:2017hua}%
  \BibitemOpen
  \bibfield  {author} {\bibinfo {author} {\bibfnamefont {J.}~\bibnamefont {Liu}}, \bibinfo {author} {\bibfnamefont {Z.-K.}\ \bibnamefont {Guo}}, \bibinfo {author} {\bibfnamefont {R.-G.}\ \bibnamefont {Cai}}, \ and\ \bibinfo {author} {\bibfnamefont {G.}~\bibnamefont {Shiu}},\ }\href {\doibase 10.1103/PhysRevLett.120.031301} {\bibfield  {journal} {\bibinfo  {journal} {Phys. Rev. Lett.}\ }\textbf {\bibinfo {volume} {120}},\ \bibinfo {pages} {031301} (\bibinfo {year} {2018})},\ \Eprint {http://arxiv.org/abs/1707.09841} {arXiv:1707.09841 [astro-ph.CO]} \BibitemShut {NoStop}%
\bibitem [{\citenamefont {Lozanov}\ and\ \citenamefont {Amin}(2019)}]{Lozanov:2019ylm}%
  \BibitemOpen
  \bibfield  {author} {\bibinfo {author} {\bibfnamefont {K.~D.}\ \bibnamefont {Lozanov}}\ and\ \bibinfo {author} {\bibfnamefont {M.~A.}\ \bibnamefont {Amin}},\ }\href {\doibase 10.1103/PhysRevD.99.123504} {\bibfield  {journal} {\bibinfo  {journal} {Phys. Rev. D}\ }\textbf {\bibinfo {volume} {99}},\ \bibinfo {pages} {123504} (\bibinfo {year} {2019})},\ \Eprint {http://arxiv.org/abs/1902.06736} {arXiv:1902.06736 [astro-ph.CO]} \BibitemShut {NoStop}%
\bibitem [{\citenamefont {Amin}\ \emph {et~al.}(2018)\citenamefont {Amin}, \citenamefont {Braden}, \citenamefont {Copeland}, \citenamefont {Giblin}, \citenamefont {Solorio}, \citenamefont {Weiner},\ and\ \citenamefont {Zhou}}]{Amin:2018xfe}%
  \BibitemOpen
  \bibfield  {author} {\bibinfo {author} {\bibfnamefont {M.~A.}\ \bibnamefont {Amin}}, \bibinfo {author} {\bibfnamefont {J.}~\bibnamefont {Braden}}, \bibinfo {author} {\bibfnamefont {E.~J.}\ \bibnamefont {Copeland}}, \bibinfo {author} {\bibfnamefont {J.~T.}\ \bibnamefont {Giblin}}, \bibinfo {author} {\bibfnamefont {C.}~\bibnamefont {Solorio}}, \bibinfo {author} {\bibfnamefont {Z.~J.}\ \bibnamefont {Weiner}}, \ and\ \bibinfo {author} {\bibfnamefont {S.-Y.}\ \bibnamefont {Zhou}},\ }\href {\doibase 10.1103/PhysRevD.98.024040} {\bibfield  {journal} {\bibinfo  {journal} {Phys. Rev. D}\ }\textbf {\bibinfo {volume} {98}},\ \bibinfo {pages} {024040} (\bibinfo {year} {2018})},\ \Eprint {http://arxiv.org/abs/1803.08047} {arXiv:1803.08047 [astro-ph.CO]} \BibitemShut {NoStop}%
\bibitem [{\citenamefont {Hiramatsu}\ \emph {et~al.}(2021)\citenamefont {Hiramatsu}, \citenamefont {Sfakianakis},\ and\ \citenamefont {Yamaguchi}}]{Hiramatsu:2020obh}%
  \BibitemOpen
  \bibfield  {author} {\bibinfo {author} {\bibfnamefont {T.}~\bibnamefont {Hiramatsu}}, \bibinfo {author} {\bibfnamefont {E.~I.}\ \bibnamefont {Sfakianakis}}, \ and\ \bibinfo {author} {\bibfnamefont {M.}~\bibnamefont {Yamaguchi}},\ }\href {\doibase 10.1007/JHEP03(2021)021} {\bibfield  {journal} {\bibinfo  {journal} {JHEP}\ }\textbf {\bibinfo {volume} {03}},\ \bibinfo {pages} {021} (\bibinfo {year} {2021})},\ \Eprint {http://arxiv.org/abs/2011.12201} {arXiv:2011.12201 [hep-ph]} \BibitemShut {NoStop}%
\bibitem [{\citenamefont {Lozanov}\ \emph {et~al.}(2023{\natexlab{a}})\citenamefont {Lozanov}, \citenamefont {Sasaki},\ and\ \citenamefont {Takhistov}}]{Lozanov:2023aez}%
  \BibitemOpen
  \bibfield  {author} {\bibinfo {author} {\bibfnamefont {K.~D.}\ \bibnamefont {Lozanov}}, \bibinfo {author} {\bibfnamefont {M.}~\bibnamefont {Sasaki}}, \ and\ \bibinfo {author} {\bibfnamefont {V.}~\bibnamefont {Takhistov}},\ }\href@noop {} {\  (\bibinfo {year} {2023}{\natexlab{a}})},\ \Eprint {http://arxiv.org/abs/2304.06709} {arXiv:2304.06709 [astro-ph.CO]} \BibitemShut {NoStop}%
\bibitem [{\citenamefont {Lozanov}\ \emph {et~al.}(2024)\citenamefont {Lozanov}, \citenamefont {Sasaki},\ and\ \citenamefont {Takhistov}}]{Lozanov:2023knf}%
  \BibitemOpen
  \bibfield  {author} {\bibinfo {author} {\bibfnamefont {K.~D.}\ \bibnamefont {Lozanov}}, \bibinfo {author} {\bibfnamefont {M.}~\bibnamefont {Sasaki}}, \ and\ \bibinfo {author} {\bibfnamefont {V.}~\bibnamefont {Takhistov}},\ }\href {\doibase 10.1016/j.physletb.2023.138392} {\bibfield  {journal} {\bibinfo  {journal} {Phys. Lett. B}\ }\textbf {\bibinfo {volume} {848}},\ \bibinfo {pages} {138392} (\bibinfo {year} {2024})},\ \Eprint {http://arxiv.org/abs/2309.14193} {arXiv:2309.14193 [astro-ph.CO]} \BibitemShut {NoStop}%
\bibitem [{\citenamefont {Lozanov}\ \emph {et~al.}(2023{\natexlab{b}})\citenamefont {Lozanov}, \citenamefont {Pi}, \citenamefont {Sasaki}, \citenamefont {Takhistov},\ and\ \citenamefont {Wang}}]{Lozanov:2023rcd}%
  \BibitemOpen
  \bibfield  {author} {\bibinfo {author} {\bibfnamefont {K.~D.}\ \bibnamefont {Lozanov}}, \bibinfo {author} {\bibfnamefont {S.}~\bibnamefont {Pi}}, \bibinfo {author} {\bibfnamefont {M.}~\bibnamefont {Sasaki}}, \bibinfo {author} {\bibfnamefont {V.}~\bibnamefont {Takhistov}}, \ and\ \bibinfo {author} {\bibfnamefont {A.}~\bibnamefont {Wang}},\ }\href@noop {} {\  (\bibinfo {year} {2023}{\natexlab{b}})},\ \Eprint {http://arxiv.org/abs/2310.03594} {arXiv:2310.03594 [astro-ph.CO]} \BibitemShut {NoStop}%
\bibitem [{\citenamefont {Lozanov}\ and\ \citenamefont {Takhistov}(2023)}]{Lozanov:2022yoy}%
  \BibitemOpen
  \bibfield  {author} {\bibinfo {author} {\bibfnamefont {K.~D.}\ \bibnamefont {Lozanov}}\ and\ \bibinfo {author} {\bibfnamefont {V.}~\bibnamefont {Takhistov}},\ }\href {\doibase 10.1103/PhysRevLett.130.181002} {\bibfield  {journal} {\bibinfo  {journal} {Phys. Rev. Lett.}\ }\textbf {\bibinfo {volume} {130}},\ \bibinfo {pages} {181002} (\bibinfo {year} {2023})},\ \Eprint {http://arxiv.org/abs/2204.07152} {arXiv:2204.07152 [astro-ph.CO]} \BibitemShut {NoStop}%
\bibitem [{\citenamefont {Liddle}\ and\ \citenamefont {Lyth}(2000)}]{Liddle:2000cg}%
  \BibitemOpen
  \bibfield  {author} {\bibinfo {author} {\bibfnamefont {A.~R.}\ \bibnamefont {Liddle}}\ and\ \bibinfo {author} {\bibfnamefont {D.~H.}\ \bibnamefont {Lyth}},\ }\href {\doibase 10.1017/CBO9781139175180} {\emph {\bibinfo {title} {{Cosmological inflation and large scale structure}}}}\ (\bibinfo {year} {2000})\BibitemShut {NoStop}%
\bibitem [{\citenamefont {Kofman}\ \emph {et~al.}(1994)\citenamefont {Kofman}, \citenamefont {Linde},\ and\ \citenamefont {Starobinsky}}]{Kofman:1994rk}%
  \BibitemOpen
  \bibfield  {author} {\bibinfo {author} {\bibfnamefont {L.}~\bibnamefont {Kofman}}, \bibinfo {author} {\bibfnamefont {A.~D.}\ \bibnamefont {Linde}}, \ and\ \bibinfo {author} {\bibfnamefont {A.~A.}\ \bibnamefont {Starobinsky}},\ }\href {\doibase 10.1103/PhysRevLett.73.3195} {\bibfield  {journal} {\bibinfo  {journal} {Phys. Rev. Lett.}\ }\textbf {\bibinfo {volume} {73}},\ \bibinfo {pages} {3195} (\bibinfo {year} {1994})},\ \Eprint {http://arxiv.org/abs/hep-th/9405187} {arXiv:hep-th/9405187} \BibitemShut {NoStop}%
\bibitem [{\citenamefont {Kofman}\ \emph {et~al.}(1997)\citenamefont {Kofman}, \citenamefont {Linde},\ and\ \citenamefont {Starobinsky}}]{Kofman:1997yn}%
  \BibitemOpen
  \bibfield  {author} {\bibinfo {author} {\bibfnamefont {L.}~\bibnamefont {Kofman}}, \bibinfo {author} {\bibfnamefont {A.~D.}\ \bibnamefont {Linde}}, \ and\ \bibinfo {author} {\bibfnamefont {A.~A.}\ \bibnamefont {Starobinsky}},\ }\href {\doibase 10.1103/PhysRevD.56.3258} {\bibfield  {journal} {\bibinfo  {journal} {Phys. Rev. D}\ }\textbf {\bibinfo {volume} {56}},\ \bibinfo {pages} {3258} (\bibinfo {year} {1997})},\ \Eprint {http://arxiv.org/abs/hep-ph/9704452} {arXiv:hep-ph/9704452} \BibitemShut {NoStop}%
\bibitem [{\citenamefont {Felder}\ \emph {et~al.}(1999)\citenamefont {Felder}, \citenamefont {Kofman},\ and\ \citenamefont {Linde}}]{Felder:1998vq}%
  \BibitemOpen
  \bibfield  {author} {\bibinfo {author} {\bibfnamefont {G.~N.}\ \bibnamefont {Felder}}, \bibinfo {author} {\bibfnamefont {L.}~\bibnamefont {Kofman}}, \ and\ \bibinfo {author} {\bibfnamefont {A.~D.}\ \bibnamefont {Linde}},\ }\href {\doibase 10.1103/PhysRevD.59.123523} {\bibfield  {journal} {\bibinfo  {journal} {Phys. Rev. D}\ }\textbf {\bibinfo {volume} {59}},\ \bibinfo {pages} {123523} (\bibinfo {year} {1999})},\ \Eprint {http://arxiv.org/abs/hep-ph/9812289} {arXiv:hep-ph/9812289} \BibitemShut {NoStop}%
\bibitem [{\citenamefont {Bezrukov}\ and\ \citenamefont {Shaposhnikov}(2008)}]{Bezrukov:2007ep}%
  \BibitemOpen
  \bibfield  {author} {\bibinfo {author} {\bibfnamefont {F.~L.}\ \bibnamefont {Bezrukov}}\ and\ \bibinfo {author} {\bibfnamefont {M.}~\bibnamefont {Shaposhnikov}},\ }\href {\doibase 10.1016/j.physletb.2007.11.072} {\bibfield  {journal} {\bibinfo  {journal} {Phys. Lett. B}\ }\textbf {\bibinfo {volume} {659}},\ \bibinfo {pages} {703} (\bibinfo {year} {2008})},\ \Eprint {http://arxiv.org/abs/0710.3755} {arXiv:0710.3755 [hep-th]} \BibitemShut {NoStop}%
\bibitem [{\citenamefont {Kawasaki}\ \emph {et~al.}(2000)\citenamefont {Kawasaki}, \citenamefont {Yamaguchi},\ and\ \citenamefont {Yanagida}}]{Kawasaki:2000yn}%
  \BibitemOpen
  \bibfield  {author} {\bibinfo {author} {\bibfnamefont {M.}~\bibnamefont {Kawasaki}}, \bibinfo {author} {\bibfnamefont {M.}~\bibnamefont {Yamaguchi}}, \ and\ \bibinfo {author} {\bibfnamefont {T.}~\bibnamefont {Yanagida}},\ }\href {\doibase 10.1103/PhysRevLett.85.3572} {\bibfield  {journal} {\bibinfo  {journal} {Phys. Rev. Lett.}\ }\textbf {\bibinfo {volume} {85}},\ \bibinfo {pages} {3572} (\bibinfo {year} {2000})},\ \Eprint {http://arxiv.org/abs/hep-ph/0004243} {arXiv:hep-ph/0004243} \BibitemShut {NoStop}%
\bibitem [{\citenamefont {Akrami}\ \emph {et~al.}(2020)\citenamefont {Akrami} \emph {et~al.}}]{Planck:2018jri}%
  \BibitemOpen
  \bibfield  {author} {\bibinfo {author} {\bibfnamefont {Y.}~\bibnamefont {Akrami}} \emph {et~al.} (\bibinfo {collaboration} {Planck}),\ }\href {\doibase 10.1051/0004-6361/201833887} {\bibfield  {journal} {\bibinfo  {journal} {Astron. Astrophys.}\ }\textbf {\bibinfo {volume} {641}},\ \bibinfo {pages} {A10} (\bibinfo {year} {2020})},\ \Eprint {http://arxiv.org/abs/1807.06211} {arXiv:1807.06211 [astro-ph.CO]} \BibitemShut {NoStop}%
\bibitem [{\citenamefont {Nakayama}\ \emph {et~al.}(2013{\natexlab{a}})\citenamefont {Nakayama}, \citenamefont {Takahashi},\ and\ \citenamefont {Yanagida}}]{Nakayama:2013jka}%
  \BibitemOpen
  \bibfield  {author} {\bibinfo {author} {\bibfnamefont {K.}~\bibnamefont {Nakayama}}, \bibinfo {author} {\bibfnamefont {F.}~\bibnamefont {Takahashi}}, \ and\ \bibinfo {author} {\bibfnamefont {T.~T.}\ \bibnamefont {Yanagida}},\ }\href {\doibase 10.1016/j.physletb.2013.06.050} {\bibfield  {journal} {\bibinfo  {journal} {Phys. Lett. B}\ }\textbf {\bibinfo {volume} {725}},\ \bibinfo {pages} {111} (\bibinfo {year} {2013}{\natexlab{a}})},\ \Eprint {http://arxiv.org/abs/1303.7315} {arXiv:1303.7315 [hep-ph]} \BibitemShut {NoStop}%
\bibitem [{\citenamefont {Nakayama}\ \emph {et~al.}(2013{\natexlab{b}})\citenamefont {Nakayama}, \citenamefont {Takahashi},\ and\ \citenamefont {Yanagida}}]{Nakayama:2013txa}%
  \BibitemOpen
  \bibfield  {author} {\bibinfo {author} {\bibfnamefont {K.}~\bibnamefont {Nakayama}}, \bibinfo {author} {\bibfnamefont {F.}~\bibnamefont {Takahashi}}, \ and\ \bibinfo {author} {\bibfnamefont {T.~T.}\ \bibnamefont {Yanagida}},\ }\href {\doibase 10.1088/1475-7516/2013/08/038} {\bibfield  {journal} {\bibinfo  {journal} {JCAP}\ }\textbf {\bibinfo {volume} {08}},\ \bibinfo {pages} {038} (\bibinfo {year} {2013}{\natexlab{b}})},\ \Eprint {http://arxiv.org/abs/1305.5099} {arXiv:1305.5099 [hep-ph]} \BibitemShut {NoStop}%
\bibitem [{\citenamefont {Kallosh}\ and\ \citenamefont {Linde}(2013)}]{Kallosh:2013hoa}%
  \BibitemOpen
  \bibfield  {author} {\bibinfo {author} {\bibfnamefont {R.}~\bibnamefont {Kallosh}}\ and\ \bibinfo {author} {\bibfnamefont {A.}~\bibnamefont {Linde}},\ }\href {\doibase 10.1088/1475-7516/2013/07/002} {\bibfield  {journal} {\bibinfo  {journal} {JCAP}\ }\textbf {\bibinfo {volume} {07}},\ \bibinfo {pages} {002} (\bibinfo {year} {2013})},\ \Eprint {http://arxiv.org/abs/1306.5220} {arXiv:1306.5220 [hep-th]} \BibitemShut {NoStop}%
\bibitem [{\citenamefont {Kallosh}\ \emph {et~al.}(2013)\citenamefont {Kallosh}, \citenamefont {Linde},\ and\ \citenamefont {Roest}}]{Kallosh:2013yoa}%
  \BibitemOpen
  \bibfield  {author} {\bibinfo {author} {\bibfnamefont {R.}~\bibnamefont {Kallosh}}, \bibinfo {author} {\bibfnamefont {A.}~\bibnamefont {Linde}}, \ and\ \bibinfo {author} {\bibfnamefont {D.}~\bibnamefont {Roest}},\ }\href {\doibase 10.1007/JHEP11(2013)198} {\bibfield  {journal} {\bibinfo  {journal} {JHEP}\ }\textbf {\bibinfo {volume} {11}},\ \bibinfo {pages} {198} (\bibinfo {year} {2013})},\ \Eprint {http://arxiv.org/abs/1311.0472} {arXiv:1311.0472 [hep-th]} \BibitemShut {NoStop}%
\bibitem [{\citenamefont {Galante}\ \emph {et~al.}(2015)\citenamefont {Galante}, \citenamefont {Kallosh}, \citenamefont {Linde},\ and\ \citenamefont {Roest}}]{Galante:2014ifa}%
  \BibitemOpen
  \bibfield  {author} {\bibinfo {author} {\bibfnamefont {M.}~\bibnamefont {Galante}}, \bibinfo {author} {\bibfnamefont {R.}~\bibnamefont {Kallosh}}, \bibinfo {author} {\bibfnamefont {A.}~\bibnamefont {Linde}}, \ and\ \bibinfo {author} {\bibfnamefont {D.}~\bibnamefont {Roest}},\ }\href {\doibase 10.1103/PhysRevLett.114.141302} {\bibfield  {journal} {\bibinfo  {journal} {Phys. Rev. Lett.}\ }\textbf {\bibinfo {volume} {114}},\ \bibinfo {pages} {141302} (\bibinfo {year} {2015})},\ \Eprint {http://arxiv.org/abs/1412.3797} {arXiv:1412.3797 [hep-th]} \BibitemShut {NoStop}%
\bibitem [{\citenamefont {Nakayama}\ \emph {et~al.}(2016)\citenamefont {Nakayama}, \citenamefont {Takahashi},\ and\ \citenamefont {Yanagida}}]{Nakayama:2016gvg}%
  \BibitemOpen
  \bibfield  {author} {\bibinfo {author} {\bibfnamefont {K.}~\bibnamefont {Nakayama}}, \bibinfo {author} {\bibfnamefont {F.}~\bibnamefont {Takahashi}}, \ and\ \bibinfo {author} {\bibfnamefont {T.~T.}\ \bibnamefont {Yanagida}},\ }\href {\doibase 10.1016/j.physletb.2016.03.051} {\bibfield  {journal} {\bibinfo  {journal} {Phys. Lett. B}\ }\textbf {\bibinfo {volume} {757}},\ \bibinfo {pages} {32} (\bibinfo {year} {2016})},\ \Eprint {http://arxiv.org/abs/1601.00192} {arXiv:1601.00192 [hep-ph]} \BibitemShut {NoStop}%
\bibitem [{\citenamefont {Bezrukov}\ \emph {et~al.}(2009)\citenamefont {Bezrukov}, \citenamefont {Gorbunov},\ and\ \citenamefont {Shaposhnikov}}]{Bezrukov:2008ut}%
  \BibitemOpen
  \bibfield  {author} {\bibinfo {author} {\bibfnamefont {F.}~\bibnamefont {Bezrukov}}, \bibinfo {author} {\bibfnamefont {D.}~\bibnamefont {Gorbunov}}, \ and\ \bibinfo {author} {\bibfnamefont {M.}~\bibnamefont {Shaposhnikov}},\ }\href {\doibase 10.1088/1475-7516/2009/06/029} {\bibfield  {journal} {\bibinfo  {journal} {JCAP}\ }\textbf {\bibinfo {volume} {06}},\ \bibinfo {pages} {029} (\bibinfo {year} {2009})},\ \Eprint {http://arxiv.org/abs/0812.3622} {arXiv:0812.3622 [hep-ph]} \BibitemShut {NoStop}%
\bibitem [{\citenamefont {Garcia-Bellido}\ \emph {et~al.}(2009)\citenamefont {Garcia-Bellido}, \citenamefont {Figueroa},\ and\ \citenamefont {Rubio}}]{Garcia-Bellido:2008ycs}%
  \BibitemOpen
  \bibfield  {author} {\bibinfo {author} {\bibfnamefont {J.}~\bibnamefont {Garcia-Bellido}}, \bibinfo {author} {\bibfnamefont {D.~G.}\ \bibnamefont {Figueroa}}, \ and\ \bibinfo {author} {\bibfnamefont {J.}~\bibnamefont {Rubio}},\ }\href {\doibase 10.1103/PhysRevD.79.063531} {\bibfield  {journal} {\bibinfo  {journal} {Phys. Rev. D}\ }\textbf {\bibinfo {volume} {79}},\ \bibinfo {pages} {063531} (\bibinfo {year} {2009})},\ \Eprint {http://arxiv.org/abs/0812.4624} {arXiv:0812.4624 [hep-ph]} \BibitemShut {NoStop}%
\bibitem [{\citenamefont {Ema}\ \emph {et~al.}(2017)\citenamefont {Ema}, \citenamefont {Jinno}, \citenamefont {Mukaida},\ and\ \citenamefont {Nakayama}}]{Ema:2016dny}%
  \BibitemOpen
  \bibfield  {author} {\bibinfo {author} {\bibfnamefont {Y.}~\bibnamefont {Ema}}, \bibinfo {author} {\bibfnamefont {R.}~\bibnamefont {Jinno}}, \bibinfo {author} {\bibfnamefont {K.}~\bibnamefont {Mukaida}}, \ and\ \bibinfo {author} {\bibfnamefont {K.}~\bibnamefont {Nakayama}},\ }\href {\doibase 10.1088/1475-7516/2017/02/045} {\bibfield  {journal} {\bibinfo  {journal} {JCAP}\ }\textbf {\bibinfo {volume} {02}},\ \bibinfo {pages} {045} (\bibinfo {year} {2017})},\ \Eprint {http://arxiv.org/abs/1609.05209} {arXiv:1609.05209 [hep-ph]} \BibitemShut {NoStop}%
\bibitem [{\citenamefont {DeCross}\ \emph {et~al.}(2018)\citenamefont {DeCross}, \citenamefont {Kaiser}, \citenamefont {Prabhu}, \citenamefont {Prescod-Weinstein},\ and\ \citenamefont {Sfakianakis}}]{DeCross:2016cbs}%
  \BibitemOpen
  \bibfield  {author} {\bibinfo {author} {\bibfnamefont {M.~P.}\ \bibnamefont {DeCross}}, \bibinfo {author} {\bibfnamefont {D.~I.}\ \bibnamefont {Kaiser}}, \bibinfo {author} {\bibfnamefont {A.}~\bibnamefont {Prabhu}}, \bibinfo {author} {\bibfnamefont {C.}~\bibnamefont {Prescod-Weinstein}}, \ and\ \bibinfo {author} {\bibfnamefont {E.~I.}\ \bibnamefont {Sfakianakis}},\ }\href {\doibase 10.1103/PhysRevD.97.023528} {\bibfield  {journal} {\bibinfo  {journal} {Phys. Rev. D}\ }\textbf {\bibinfo {volume} {97}},\ \bibinfo {pages} {023528} (\bibinfo {year} {2018})},\ \Eprint {http://arxiv.org/abs/1610.08916} {arXiv:1610.08916 [astro-ph.CO]} \BibitemShut {NoStop}%
\bibitem [{\citenamefont {Sfakianakis}\ and\ \citenamefont {van~de Vis}(2019)}]{Sfakianakis:2018lzf}%
  \BibitemOpen
  \bibfield  {author} {\bibinfo {author} {\bibfnamefont {E.~I.}\ \bibnamefont {Sfakianakis}}\ and\ \bibinfo {author} {\bibfnamefont {J.}~\bibnamefont {van~de Vis}},\ }\href {\doibase 10.1103/PhysRevD.99.083519} {\bibfield  {journal} {\bibinfo  {journal} {Phys. Rev. D}\ }\textbf {\bibinfo {volume} {99}},\ \bibinfo {pages} {083519} (\bibinfo {year} {2019})},\ \Eprint {http://arxiv.org/abs/1810.01304} {arXiv:1810.01304 [hep-ph]} \BibitemShut {NoStop}%
\bibitem [{\citenamefont {Ema}\ \emph {et~al.}(2021)\citenamefont {Ema}, \citenamefont {Jinno}, \citenamefont {Nakayama},\ and\ \citenamefont {van~de Vis}}]{Ema:2021xhq}%
  \BibitemOpen
  \bibfield  {author} {\bibinfo {author} {\bibfnamefont {Y.}~\bibnamefont {Ema}}, \bibinfo {author} {\bibfnamefont {R.}~\bibnamefont {Jinno}}, \bibinfo {author} {\bibfnamefont {K.}~\bibnamefont {Nakayama}}, \ and\ \bibinfo {author} {\bibfnamefont {J.}~\bibnamefont {van~de Vis}},\ }\href {\doibase 10.1103/PhysRevD.103.103536} {\bibfield  {journal} {\bibinfo  {journal} {Phys. Rev. D}\ }\textbf {\bibinfo {volume} {103}},\ \bibinfo {pages} {103536} (\bibinfo {year} {2021})},\ \Eprint {http://arxiv.org/abs/2102.12501} {arXiv:2102.12501 [hep-ph]} \BibitemShut {NoStop}%
\bibitem [{\citenamefont {Bauer}\ and\ \citenamefont {Demir}(2008)}]{Bauer:2008zj}%
  \BibitemOpen
  \bibfield  {author} {\bibinfo {author} {\bibfnamefont {F.}~\bibnamefont {Bauer}}\ and\ \bibinfo {author} {\bibfnamefont {D.~A.}\ \bibnamefont {Demir}},\ }\href {\doibase 10.1016/j.physletb.2008.06.014} {\bibfield  {journal} {\bibinfo  {journal} {Phys. Lett. B}\ }\textbf {\bibinfo {volume} {665}},\ \bibinfo {pages} {222} (\bibinfo {year} {2008})},\ \Eprint {http://arxiv.org/abs/0803.2664} {arXiv:0803.2664 [hep-ph]} \BibitemShut {NoStop}%
\bibitem [{\citenamefont {Bauer}\ and\ \citenamefont {Demir}(2011)}]{Bauer:2010jg}%
  \BibitemOpen
  \bibfield  {author} {\bibinfo {author} {\bibfnamefont {F.}~\bibnamefont {Bauer}}\ and\ \bibinfo {author} {\bibfnamefont {D.~A.}\ \bibnamefont {Demir}},\ }\href {\doibase 10.1016/j.physletb.2011.03.042} {\bibfield  {journal} {\bibinfo  {journal} {Phys. Lett. B}\ }\textbf {\bibinfo {volume} {698}},\ \bibinfo {pages} {425} (\bibinfo {year} {2011})},\ \Eprint {http://arxiv.org/abs/1012.2900} {arXiv:1012.2900 [hep-ph]} \BibitemShut {NoStop}%
\bibitem [{\citenamefont {Rubio}\ and\ \citenamefont {Tomberg}(2019)}]{Rubio:2019ypq}%
  \BibitemOpen
  \bibfield  {author} {\bibinfo {author} {\bibfnamefont {J.}~\bibnamefont {Rubio}}\ and\ \bibinfo {author} {\bibfnamefont {E.~S.}\ \bibnamefont {Tomberg}},\ }\href {\doibase 10.1088/1475-7516/2019/04/021} {\bibfield  {journal} {\bibinfo  {journal} {JCAP}\ }\textbf {\bibinfo {volume} {04}},\ \bibinfo {pages} {021} (\bibinfo {year} {2019})},\ \Eprint {http://arxiv.org/abs/1902.10148} {arXiv:1902.10148 [hep-ph]} \BibitemShut {NoStop}%
\bibitem [{\citenamefont {Murayama}\ \emph {et~al.}(2014)\citenamefont {Murayama}, \citenamefont {Nakayama}, \citenamefont {Takahashi},\ and\ \citenamefont {Yanagida}}]{Murayama:2014saa}%
  \BibitemOpen
  \bibfield  {author} {\bibinfo {author} {\bibfnamefont {H.}~\bibnamefont {Murayama}}, \bibinfo {author} {\bibfnamefont {K.}~\bibnamefont {Nakayama}}, \bibinfo {author} {\bibfnamefont {F.}~\bibnamefont {Takahashi}}, \ and\ \bibinfo {author} {\bibfnamefont {T.~T.}\ \bibnamefont {Yanagida}},\ }\href {\doibase 10.1016/j.physletb.2014.09.045} {\bibfield  {journal} {\bibinfo  {journal} {Phys. Lett. B}\ }\textbf {\bibinfo {volume} {738}},\ \bibinfo {pages} {196} (\bibinfo {year} {2014})},\ \Eprint {http://arxiv.org/abs/1404.3857} {arXiv:1404.3857 [hep-ph]} \BibitemShut {NoStop}%
\bibitem [{\citenamefont {Evans}\ \emph {et~al.}(2015)\citenamefont {Evans}, \citenamefont {Gherghetta},\ and\ \citenamefont {Peloso}}]{Evans:2015mta}%
  \BibitemOpen
  \bibfield  {author} {\bibinfo {author} {\bibfnamefont {J.~L.}\ \bibnamefont {Evans}}, \bibinfo {author} {\bibfnamefont {T.}~\bibnamefont {Gherghetta}}, \ and\ \bibinfo {author} {\bibfnamefont {M.}~\bibnamefont {Peloso}},\ }\href {\doibase 10.1103/PhysRevD.92.021303} {\bibfield  {journal} {\bibinfo  {journal} {Phys. Rev. D}\ }\textbf {\bibinfo {volume} {92}},\ \bibinfo {pages} {021303} (\bibinfo {year} {2015})},\ \Eprint {http://arxiv.org/abs/1501.06560} {arXiv:1501.06560 [hep-ph]} \BibitemShut {NoStop}%
\bibitem [{\citenamefont {Kallosh}\ \emph {et~al.}(2016)\citenamefont {Kallosh}, \citenamefont {Linde}, \citenamefont {Roest},\ and\ \citenamefont {Wrase}}]{Kallosh:2016sej}%
  \BibitemOpen
  \bibfield  {author} {\bibinfo {author} {\bibfnamefont {R.}~\bibnamefont {Kallosh}}, \bibinfo {author} {\bibfnamefont {A.}~\bibnamefont {Linde}}, \bibinfo {author} {\bibfnamefont {D.}~\bibnamefont {Roest}}, \ and\ \bibinfo {author} {\bibfnamefont {T.}~\bibnamefont {Wrase}},\ }\href {\doibase 10.1088/1475-7516/2016/11/046} {\bibfield  {journal} {\bibinfo  {journal} {JCAP}\ }\textbf {\bibinfo {volume} {11}},\ \bibinfo {pages} {046} (\bibinfo {year} {2016})},\ \Eprint {http://arxiv.org/abs/1607.08854} {arXiv:1607.08854 [hep-th]} \BibitemShut {NoStop}%
\bibitem [{\citenamefont {Minkowski}(1977)}]{Minkowski:1977sc}%
  \BibitemOpen
  \bibfield  {author} {\bibinfo {author} {\bibfnamefont {P.}~\bibnamefont {Minkowski}},\ }\href {\doibase 10.1016/0370-2693(77)90435-X} {\bibfield  {journal} {\bibinfo  {journal} {Phys. Lett. B}\ }\textbf {\bibinfo {volume} {67}},\ \bibinfo {pages} {421} (\bibinfo {year} {1977})}\BibitemShut {NoStop}%
\bibitem [{\citenamefont {Yanagida}(1979)}]{Yanagida:1979as}%
  \BibitemOpen
  \bibfield  {author} {\bibinfo {author} {\bibfnamefont {T.}~\bibnamefont {Yanagida}},\ }\href@noop {} {\bibfield  {journal} {\bibinfo  {journal} {Conf. Proc. C}\ }\textbf {\bibinfo {volume} {7902131}},\ \bibinfo {pages} {95} (\bibinfo {year} {1979})}\BibitemShut {NoStop}%
\bibitem [{\citenamefont {Gell-Mann}\ \emph {et~al.}(1979)\citenamefont {Gell-Mann}, \citenamefont {Ramond},\ and\ \citenamefont {Slansky}}]{Gell-Mann:1979vob}%
  \BibitemOpen
  \bibfield  {author} {\bibinfo {author} {\bibfnamefont {M.}~\bibnamefont {Gell-Mann}}, \bibinfo {author} {\bibfnamefont {P.}~\bibnamefont {Ramond}}, \ and\ \bibinfo {author} {\bibfnamefont {R.}~\bibnamefont {Slansky}},\ }\href@noop {} {\bibfield  {journal} {\bibinfo  {journal} {Conf. Proc. C}\ }\textbf {\bibinfo {volume} {790927}},\ \bibinfo {pages} {315} (\bibinfo {year} {1979})},\ \Eprint {http://arxiv.org/abs/1306.4669} {arXiv:1306.4669 [hep-th]} \BibitemShut {NoStop}%
\bibitem [{\citenamefont {Nakayama}\ \emph {et~al.}(2017)\citenamefont {Nakayama}, \citenamefont {Takahashi},\ and\ \citenamefont {Yanagida}}]{Nakayama:2017cij}%
  \BibitemOpen
  \bibfield  {author} {\bibinfo {author} {\bibfnamefont {K.}~\bibnamefont {Nakayama}}, \bibinfo {author} {\bibfnamefont {F.}~\bibnamefont {Takahashi}}, \ and\ \bibinfo {author} {\bibfnamefont {T.~T.}\ \bibnamefont {Yanagida}},\ }\href {\doibase 10.1016/j.physletb.2017.08.024} {\bibfield  {journal} {\bibinfo  {journal} {Phys. Lett. B}\ }\textbf {\bibinfo {volume} {773}},\ \bibinfo {pages} {179} (\bibinfo {year} {2017})},\ \Eprint {http://arxiv.org/abs/1705.04796} {arXiv:1705.04796 [hep-ph]} \BibitemShut {NoStop}%
\bibitem [{\citenamefont {Fukugita}\ and\ \citenamefont {Yanagida}(1986)}]{Fukugita:1986hr}%
  \BibitemOpen
  \bibfield  {author} {\bibinfo {author} {\bibfnamefont {M.}~\bibnamefont {Fukugita}}\ and\ \bibinfo {author} {\bibfnamefont {T.}~\bibnamefont {Yanagida}},\ }\href {\doibase 10.1016/0370-2693(86)91126-3} {\bibfield  {journal} {\bibinfo  {journal} {Phys. Lett. B}\ }\textbf {\bibinfo {volume} {174}},\ \bibinfo {pages} {45} (\bibinfo {year} {1986})}\BibitemShut {NoStop}%
\bibitem [{\citenamefont {Abbott}\ \emph {et~al.}(2016{\natexlab{b}})\citenamefont {Abbott} \emph {et~al.}}]{LIGOScientific:2016fpe}%
  \BibitemOpen
  \bibfield  {author} {\bibinfo {author} {\bibfnamefont {B.~P.}\ \bibnamefont {Abbott}} \emph {et~al.} (\bibinfo {collaboration} {LIGO Scientific, Virgo}),\ }\href {\doibase 10.1103/PhysRevLett.116.131102} {\bibfield  {journal} {\bibinfo  {journal} {Phys. Rev. Lett.}\ }\textbf {\bibinfo {volume} {116}},\ \bibinfo {pages} {131102} (\bibinfo {year} {2016}{\natexlab{b}})},\ \Eprint {http://arxiv.org/abs/1602.03847} {arXiv:1602.03847 [gr-qc]} \BibitemShut {NoStop}%
\bibitem [{\citenamefont {Amaro-Seoane}\ \emph {et~al.}(2017)\citenamefont {Amaro-Seoane} \emph {et~al.}}]{LISA:2017pwj}%
  \BibitemOpen
  \bibfield  {author} {\bibinfo {author} {\bibfnamefont {P.}~\bibnamefont {Amaro-Seoane}} \emph {et~al.} (\bibinfo {collaboration} {LISA}),\ }\href@noop {} {\  (\bibinfo {year} {2017})},\ \Eprint {http://arxiv.org/abs/1702.00786} {arXiv:1702.00786 [astro-ph.IM]} \BibitemShut {NoStop}%
\bibitem [{\citenamefont {Reitze}\ \emph {et~al.}(2019)\citenamefont {Reitze} \emph {et~al.}}]{Reitze:2019iox}%
  \BibitemOpen
  \bibfield  {author} {\bibinfo {author} {\bibfnamefont {D.}~\bibnamefont {Reitze}} \emph {et~al.},\ }\href@noop {} {\bibfield  {journal} {\bibinfo  {journal} {Bull. Am. Astron. Soc.}\ }\textbf {\bibinfo {volume} {51}},\ \bibinfo {pages} {035} (\bibinfo {year} {2019})},\ \Eprint {http://arxiv.org/abs/1907.04833} {arXiv:1907.04833 [astro-ph.IM]} \BibitemShut {NoStop}%
\bibitem [{\citenamefont {Punturo}\ \emph {et~al.}(2010)\citenamefont {Punturo} \emph {et~al.}}]{Punturo:2010zz}%
  \BibitemOpen
  \bibfield  {author} {\bibinfo {author} {\bibfnamefont {M.}~\bibnamefont {Punturo}} \emph {et~al.},\ }\href {\doibase 10.1088/0264-9381/27/19/194002} {\bibfield  {journal} {\bibinfo  {journal} {Class. Quant. Grav.}\ }\textbf {\bibinfo {volume} {27}},\ \bibinfo {pages} {194002} (\bibinfo {year} {2010})}\BibitemShut {NoStop}%
\bibitem [{\citenamefont {Harry}\ \emph {et~al.}(2006)\citenamefont {Harry}, \citenamefont {Fritschel}, \citenamefont {Shaddock}, \citenamefont {Folkner},\ and\ \citenamefont {Phinney}}]{Harry:2006fi}%
  \BibitemOpen
  \bibfield  {author} {\bibinfo {author} {\bibfnamefont {G.~M.}\ \bibnamefont {Harry}}, \bibinfo {author} {\bibfnamefont {P.}~\bibnamefont {Fritschel}}, \bibinfo {author} {\bibfnamefont {D.~A.}\ \bibnamefont {Shaddock}}, \bibinfo {author} {\bibfnamefont {W.}~\bibnamefont {Folkner}}, \ and\ \bibinfo {author} {\bibfnamefont {E.~S.}\ \bibnamefont {Phinney}},\ }\href {\doibase 10.1088/0264-9381/23/15/008} {\bibfield  {journal} {\bibinfo  {journal} {Class. Quant. Grav.}\ }\textbf {\bibinfo {volume} {23}},\ \bibinfo {pages} {4887} (\bibinfo {year} {2006})},\ \bibinfo {note} {[Erratum: Class.Quant.Grav. 23, 7361 (2006)]}\BibitemShut {NoStop}%
\bibitem [{\citenamefont {Seto}\ \emph {et~al.}(2001)\citenamefont {Seto}, \citenamefont {Kawamura},\ and\ \citenamefont {Nakamura}}]{Seto:2001qf}%
  \BibitemOpen
  \bibfield  {author} {\bibinfo {author} {\bibfnamefont {N.}~\bibnamefont {Seto}}, \bibinfo {author} {\bibfnamefont {S.}~\bibnamefont {Kawamura}}, \ and\ \bibinfo {author} {\bibfnamefont {T.}~\bibnamefont {Nakamura}},\ }\href {\doibase 10.1103/PhysRevLett.87.221103} {\bibfield  {journal} {\bibinfo  {journal} {Phys. Rev. Lett.}\ }\textbf {\bibinfo {volume} {87}},\ \bibinfo {pages} {221103} (\bibinfo {year} {2001})},\ \Eprint {http://arxiv.org/abs/astro-ph/0108011} {arXiv:astro-ph/0108011} \BibitemShut {NoStop}%
\bibitem [{\citenamefont {Herman}\ \emph {et~al.}(2021)\citenamefont {Herman}, \citenamefont {F\"uzfa}, \citenamefont {Lehoucq},\ and\ \citenamefont {Clesse}}]{Herman:2020wao}%
  \BibitemOpen
  \bibfield  {author} {\bibinfo {author} {\bibfnamefont {N.}~\bibnamefont {Herman}}, \bibinfo {author} {\bibfnamefont {A.}~\bibnamefont {F\"uzfa}}, \bibinfo {author} {\bibfnamefont {L.}~\bibnamefont {Lehoucq}}, \ and\ \bibinfo {author} {\bibfnamefont {S.}~\bibnamefont {Clesse}},\ }\href {\doibase 10.1103/PhysRevD.104.023524} {\bibfield  {journal} {\bibinfo  {journal} {Phys. Rev. D}\ }\textbf {\bibinfo {volume} {104}},\ \bibinfo {pages} {023524} (\bibinfo {year} {2021})},\ \Eprint {http://arxiv.org/abs/2012.12189} {arXiv:2012.12189 [gr-qc]} \BibitemShut {NoStop}%
\bibitem [{\citenamefont {Herman}\ \emph {et~al.}(2023)\citenamefont {Herman}, \citenamefont {Lehoucq},\ and\ \citenamefont {F\'{u}zfa}}]{Herman:2022fau}%
  \BibitemOpen
  \bibfield  {author} {\bibinfo {author} {\bibfnamefont {N.}~\bibnamefont {Herman}}, \bibinfo {author} {\bibfnamefont {L.}~\bibnamefont {Lehoucq}}, \ and\ \bibinfo {author} {\bibfnamefont {A.}~\bibnamefont {F\'{u}zfa}},\ }\href {\doibase 10.1103/PhysRevD.108.124009} {\bibfield  {journal} {\bibinfo  {journal} {Phys. Rev. D}\ }\textbf {\bibinfo {volume} {108}},\ \bibinfo {pages} {124009} (\bibinfo {year} {2023})},\ \Eprint {http://arxiv.org/abs/2203.15668} {arXiv:2203.15668 [gr-qc]} \BibitemShut {NoStop}%
\bibitem [{\citenamefont {Khlebnikov}\ and\ \citenamefont {Tkachev}(1997)}]{Khlebnikov:1997di}%
  \BibitemOpen
  \bibfield  {author} {\bibinfo {author} {\bibfnamefont {S.~Y.}\ \bibnamefont {Khlebnikov}}\ and\ \bibinfo {author} {\bibfnamefont {I.~I.}\ \bibnamefont {Tkachev}},\ }\href {\doibase 10.1103/PhysRevD.56.653} {\bibfield  {journal} {\bibinfo  {journal} {Phys. Rev. D}\ }\textbf {\bibinfo {volume} {56}},\ \bibinfo {pages} {653} (\bibinfo {year} {1997})},\ \Eprint {http://arxiv.org/abs/hep-ph/9701423} {arXiv:hep-ph/9701423} \BibitemShut {NoStop}%
\bibitem [{\citenamefont {Easther}\ and\ \citenamefont {Lim}(2006)}]{Easther:2006gt}%
  \BibitemOpen
  \bibfield  {author} {\bibinfo {author} {\bibfnamefont {R.}~\bibnamefont {Easther}}\ and\ \bibinfo {author} {\bibfnamefont {E.~A.}\ \bibnamefont {Lim}},\ }\href {\doibase 10.1088/1475-7516/2006/04/010} {\bibfield  {journal} {\bibinfo  {journal} {JCAP}\ }\textbf {\bibinfo {volume} {04}},\ \bibinfo {pages} {010} (\bibinfo {year} {2006})},\ \Eprint {http://arxiv.org/abs/astro-ph/0601617} {arXiv:astro-ph/0601617} \BibitemShut {NoStop}%
\bibitem [{\citenamefont {Easther}\ \emph {et~al.}(2007)\citenamefont {Easther}, \citenamefont {Giblin},\ and\ \citenamefont {Lim}}]{Easther:2006vd}%
  \BibitemOpen
  \bibfield  {author} {\bibinfo {author} {\bibfnamefont {R.}~\bibnamefont {Easther}}, \bibinfo {author} {\bibfnamefont {J.~T.}\ \bibnamefont {Giblin}, \bibfnamefont {Jr.}}, \ and\ \bibinfo {author} {\bibfnamefont {E.~A.}\ \bibnamefont {Lim}},\ }\href {\doibase 10.1103/PhysRevLett.99.221301} {\bibfield  {journal} {\bibinfo  {journal} {Phys. Rev. Lett.}\ }\textbf {\bibinfo {volume} {99}},\ \bibinfo {pages} {221301} (\bibinfo {year} {2007})},\ \Eprint {http://arxiv.org/abs/astro-ph/0612294} {arXiv:astro-ph/0612294} \BibitemShut {NoStop}%
\bibitem [{\citenamefont {Garcia-Bellido}\ \emph {et~al.}(2008)\citenamefont {Garcia-Bellido}, \citenamefont {Figueroa},\ and\ \citenamefont {Sastre}}]{Garcia-Bellido:2007fiu}%
  \BibitemOpen
  \bibfield  {author} {\bibinfo {author} {\bibfnamefont {J.}~\bibnamefont {Garcia-Bellido}}, \bibinfo {author} {\bibfnamefont {D.~G.}\ \bibnamefont {Figueroa}}, \ and\ \bibinfo {author} {\bibfnamefont {A.}~\bibnamefont {Sastre}},\ }\href {\doibase 10.1103/PhysRevD.77.043517} {\bibfield  {journal} {\bibinfo  {journal} {Phys. Rev. D}\ }\textbf {\bibinfo {volume} {77}},\ \bibinfo {pages} {043517} (\bibinfo {year} {2008})},\ \Eprint {http://arxiv.org/abs/0707.0839} {arXiv:0707.0839 [hep-ph]} \BibitemShut {NoStop}%
\bibitem [{\citenamefont {Dufaux}\ \emph {et~al.}(2007)\citenamefont {Dufaux}, \citenamefont {Bergman}, \citenamefont {Felder}, \citenamefont {Kofman},\ and\ \citenamefont {Uzan}}]{Dufaux:2007pt}%
  \BibitemOpen
  \bibfield  {author} {\bibinfo {author} {\bibfnamefont {J.~F.}\ \bibnamefont {Dufaux}}, \bibinfo {author} {\bibfnamefont {A.}~\bibnamefont {Bergman}}, \bibinfo {author} {\bibfnamefont {G.~N.}\ \bibnamefont {Felder}}, \bibinfo {author} {\bibfnamefont {L.}~\bibnamefont {Kofman}}, \ and\ \bibinfo {author} {\bibfnamefont {J.-P.}\ \bibnamefont {Uzan}},\ }\href {\doibase 10.1103/PhysRevD.76.123517} {\bibfield  {journal} {\bibinfo  {journal} {Phys. Rev. D}\ }\textbf {\bibinfo {volume} {76}},\ \bibinfo {pages} {123517} (\bibinfo {year} {2007})},\ \Eprint {http://arxiv.org/abs/0707.0875} {arXiv:0707.0875 [astro-ph]} \BibitemShut {NoStop}%
\bibitem [{\citenamefont {Figueroa}\ \emph {et~al.}(2011)\citenamefont {Figueroa}, \citenamefont {Garcia-Bellido},\ and\ \citenamefont {Rajantie}}]{Figueroa:2011ye}%
  \BibitemOpen
  \bibfield  {author} {\bibinfo {author} {\bibfnamefont {D.~G.}\ \bibnamefont {Figueroa}}, \bibinfo {author} {\bibfnamefont {J.}~\bibnamefont {Garcia-Bellido}}, \ and\ \bibinfo {author} {\bibfnamefont {A.}~\bibnamefont {Rajantie}},\ }\href {\doibase 10.1088/1475-7516/2011/11/015} {\bibfield  {journal} {\bibinfo  {journal} {JCAP}\ }\textbf {\bibinfo {volume} {11}},\ \bibinfo {pages} {015} (\bibinfo {year} {2011})},\ \Eprint {http://arxiv.org/abs/1110.0337} {arXiv:1110.0337 [astro-ph.CO]} \BibitemShut {NoStop}%
\bibitem [{\citenamefont {Repond}\ and\ \citenamefont {Rubio}(2016)}]{Repond:2016sol}%
  \BibitemOpen
  \bibfield  {author} {\bibinfo {author} {\bibfnamefont {J.}~\bibnamefont {Repond}}\ and\ \bibinfo {author} {\bibfnamefont {J.}~\bibnamefont {Rubio}},\ }\href {\doibase 10.1088/1475-7516/2016/07/043} {\bibfield  {journal} {\bibinfo  {journal} {JCAP}\ }\textbf {\bibinfo {volume} {07}},\ \bibinfo {pages} {043} (\bibinfo {year} {2016})},\ \Eprint {http://arxiv.org/abs/1604.08238} {arXiv:1604.08238 [astro-ph.CO]} \BibitemShut {NoStop}%
\bibitem [{\citenamefont {Figueroa}\ and\ \citenamefont {Torrenti}(2017)}]{Figueroa:2017vfa}%
  \BibitemOpen
  \bibfield  {author} {\bibinfo {author} {\bibfnamefont {D.~G.}\ \bibnamefont {Figueroa}}\ and\ \bibinfo {author} {\bibfnamefont {F.}~\bibnamefont {Torrenti}},\ }\href {\doibase 10.1088/1475-7516/2017/10/057} {\bibfield  {journal} {\bibinfo  {journal} {JCAP}\ }\textbf {\bibinfo {volume} {10}},\ \bibinfo {pages} {057} (\bibinfo {year} {2017})},\ \Eprint {http://arxiv.org/abs/1707.04533} {arXiv:1707.04533 [astro-ph.CO]} \BibitemShut {NoStop}%
\bibitem [{\citenamefont {Figueroa}\ \emph {et~al.}(2022)\citenamefont {Figueroa}, \citenamefont {Florio}, \citenamefont {Loayza},\ and\ \citenamefont {Pieroni}}]{Figueroa:2022iho}%
  \BibitemOpen
  \bibfield  {author} {\bibinfo {author} {\bibfnamefont {D.~G.}\ \bibnamefont {Figueroa}}, \bibinfo {author} {\bibfnamefont {A.}~\bibnamefont {Florio}}, \bibinfo {author} {\bibfnamefont {N.}~\bibnamefont {Loayza}}, \ and\ \bibinfo {author} {\bibfnamefont {M.}~\bibnamefont {Pieroni}},\ }\href {\doibase 10.1103/PhysRevD.106.063522} {\bibfield  {journal} {\bibinfo  {journal} {Phys. Rev. D}\ }\textbf {\bibinfo {volume} {106}},\ \bibinfo {pages} {063522} (\bibinfo {year} {2022})},\ \Eprint {http://arxiv.org/abs/2202.05805} {arXiv:2202.05805 [astro-ph.CO]} \BibitemShut {NoStop}%
\end{thebibliography}%

\end{document}